\documentclass[letterpaper,titlepage,11pt]{article}

\pdfoutput=1

\usepackage[a4paper]{geometry}
\usepackage{amsthm,amscd,bm}
\usepackage{tensor}
\usepackage{mathtools}
\usepackage{epsfig}
\usepackage{graphicx}
\usepackage{enumerate}
\usepackage[english]{babel}
\usepackage{comment}
\usepackage[usenames,dvipsnames]{xcolor}
\usepackage{cite,enumerate,booktabs,float}
\usepackage[affil-it]{authblk}

\usepackage{caption}
\usepackage{subcaption}

\def\be{\begin{equation}}
\def\ee{\end{equation}}
\def\bea{\begin{eqnarray}}
\def\eea{\end{eqnarray}}

\usepackage{tikz}
\usetikzlibrary{cd}
\usetikzlibrary{decorations.pathmorphing}
\usetikzlibrary{arrows.meta}
\tikzset{>={Stealth[width=3mm,length=3mm]}}

\def\half{{1\over 2}}

\usepackage[T1]{fontenc}
\usepackage[utf8]{inputenc}
\usepackage{lmodern}

\usepackage{amssymb,amsmath,amsfonts}
\usepackage{epsfig}
\usepackage{graphicx}
\usepackage{epstopdf}
\usepackage{caption}
\usepackage{subcaption}
\usepackage{amscd}
\usepackage{stmaryrd}
\usepackage{amsthm}
\usepackage{latexsym}
\usepackage{amsbsy}
\usepackage[english]{babel}
\usepackage{psfrag}
\usepackage{tabularx}
\usepackage{cite}

\usepackage[pdftex]{hyperref}
\hypersetup{
    colorlinks, linkcolor={red},
    citecolor={magenta}
}

\allowdisplaybreaks[1]

\def\clock{{\count0=\time
           \divide\count0 60
           \ifnum\count0<10 0\fi\the\count0
           \multiply\count0 -60 \advance\count0 \time
           :\ifnum\count0<10 0\fi \the\count0
         }}
\newcommand{\timestamp}{{\small\vbox{\hbox{\tt\jobname.tex}
\hbox{\the\day/\the\month/\the\year, \clock}}}}

\def\nn{\nonumber}

\begin{document}

\numberwithin{equation}{section}

\begin{titlepage}
\rightline{\vbox{   \phantom{ghost} }}

\begin{flushright}
YITP-19-47
\end{flushright}

 \vskip 1.8 cm
\begin{center}
{\Large \bf \boldmath
Information Transfer and Black Hole Evaporation via Traversable BTZ Wormholes}
\end{center}
\vskip .5cm

\title{}
\date{\today}
\author{Shinji Hirano, Yang Lei, Sam van Leuven}

\centerline{\large {{\bf Shinji Hirano$^{* \dagger}$\footnote{
	e-mail:
	\href{mailto:shinji.hirano@wits.ac.za}{shinji.hirano@wits.ac.za}}, Yang Lei$^*$\footnote{
	e-mail:
	\href{mailto:leiy890712@gmail.com}{leiy890712@gmail.com}}, Sam van Leuven$^{*}$\footnote{
	e-mail:
	\href{mailto:sam.vanleuven@wits.ac.za}{sam.vanleuven@wits.ac.za}}}}}
	
	\setcounter{footnote}{0}

\vskip .8cm

\begin{center}

\sl $^*$ 
School of Physics and Mandelstam Institute for Theoretical Physics\\
DST-NRF Centre of Excellence in Mathematical and Statistical Sciences (CoE-MaSS)\\
National Institute for Theoretical Physics\\
University of the Witwatersrand, WITS 2050, Johannesburg, South Africa, \\
 $^{\dagger}$Center for Gravitational Physics,
Yukawa Institute for Theoretical Physics\\
Kyoto University, Kyoto 606-8502, Japan

\end{center}

\vskip 1.3cm \centerline{\bf Abstract} \vskip 0.2cm \noindent

We study traversable wormholes by considering the duality between BTZ black holes and two-dimensional conformal field theory on the thermofield double state. 
The BTZ black holes can be rendered traversable by a negative energy shock wave. 
Following Gao, Jafferis and Wall \cite{Gao:2016bin}, we show that the negative energy shock wave is dual to the infinite boost limit of a specific double trace deformation which couples the left and right CFTs. 
We spell out the mechanism of information transfer through traversable BTZ wormholes, treating the backreaction of the message as a positive energy shockwave. 
The corresponding spacetime is that of colliding spherical shells in the BTZ black hole, which we explicitly construct. 
This construction allows us to obtain a bound on the amount of information that can be sent through the wormhole, which is consistent with previous work in the context of nearly $AdS_2$ gravity \cite{Maldacena:2017axo}. 
Consequently, we define a notion of traversibility of the wormhole and study it in the context of a multiple shock geometry.
We argue that the time-dependence of traversibility in this geometry can be connected to certain aspects of the black hole evaporation process, such as the second half of the Page curve.
Finally, we examine the claim that traversable wormholes are fast decoders. 
We find evidence for this by computing the scrambling time in the shockwave background and showing that it is delayed by the presence of the negative energy shock wave.

\end{titlepage}

\newpage

{ \hypersetup{linkcolor=black}  \tableofcontents}

\section{Introduction}\label{sec:intro}

One of the most tantalizing problems in theoretical physics is the black hole information paradox or its more recent incarnation in the form of the firewall paradox \cite{Hawking:1976ra,Almheiri:2012rt,Almheiri:2013hfa}.
An important ingredient in both paradoxes concerns unitarity of the black hole S-matrix, which describes the evolution of infalling matter into outgoing Hawking radiation.
In particular, the firewall paradox contends that unitarity of the S-matrix cannot be consistent with smoothness of the horizon if the effective dynamics on the horizon is to be described by local quantum field theory on curved spacetime.\footnote{Some of the works devoted to resolving the paradox and relevant for the present work are given by \cite{Maldacena:2013xja,Papadodimas:2012aq,Papadodimas:2013jku,Yoshida:2019qqw,Penington:2019npb}.}

\begin{figure}[t]
	\centering \includegraphics[height=2in]{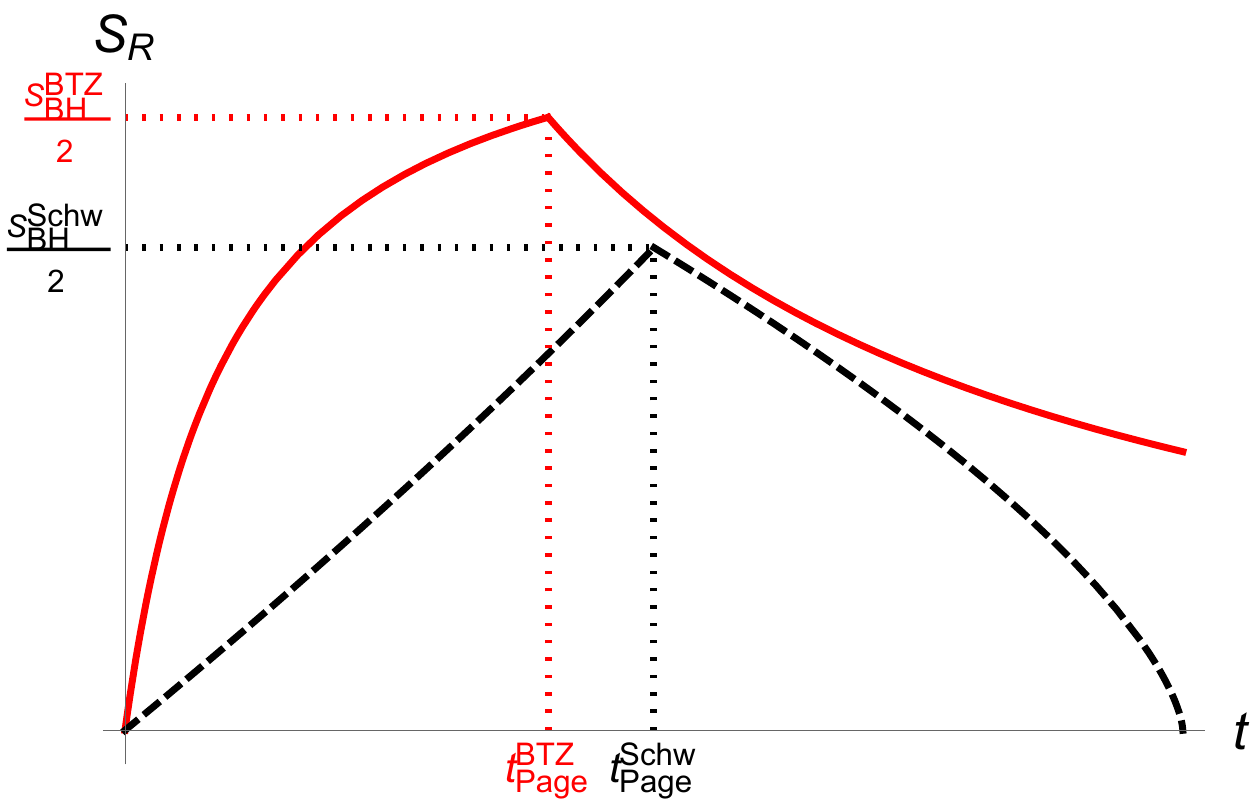} 
	\caption{\small\it The Page curves for a four-dimensional Schwarzschild black hole (black dashed) and a three-dimensional BTZ black hole (solid red). They show the time dependence of the entanglement entropy of the radiation subsystem as a function of time. The Page time is defined as the time at which the curve reaches its maximum, where the entanglement entropy equals approximately half of the Bekenstein-Hawking entropy of the original black hole. See \cite{Page:1993wv,Page:2013dx} for more details.}  \label{fig:PageCurves-intro}
\end{figure}

A full understanding of the evaporation process is still out of reach, but important progress has been made.
An apparently fruitful and widely practised way of addressing the issue is to assume unitarity of the evaporation process and study the consequences.
A seminal contribution in this regard was made by Page \cite{Page:1993wv,Page:2013dx}.
He studied the evaporating black hole as a bipartite system, consisting of the remaining black hole and the already emitted Hawking radiation.
The full system, under the assumption of a unitary evaporation process, is described by a pure state at any stage in time. 
Page's idea was to study the time dependence of the entanglement entropy of the reduced density matrix associated to the radiation subsystem, denoted by $S_R$.
Using quantum information theoretic arguments, he could determine the approximate behaviour of $S_R$ at early and late times in the evaporation process.
This results in the Page curve, as illustrated in Figure \ref{fig:PageCurves-intro}.
The maximum of this curve sits at the Page time, which can roughly be thought of as the time at which the black hole is halfway through its evaporation.
It is a turn-over point in the sense that the entanglement entropy of the radiation has become of the same order as the remaining black hole entropy. 
Up until the Page time, the reduced density matrix of the radiation is very nearly maximally mixed, i.e. it contains no information.
After the Page time, however, the drop in entanglement entropy reflects the fact that the radiation subsystem slowly purifies such that when the black hole has completely evaporated the radiation is described by a pure state.
In other words, at this stage information is slowly leaking out of the black hole.
One can argue, again using quantum information theoretic arguments, that this information cannot be accessed in any small subsystem of the Hawking radiation.
Instead, it should reside in correlations between large numbers of Hawking particles.

Another important development, based on the unitarity assumption, was made by Hayden and Preskill \cite{Hayden:2007cs}.
They present a thought experiment that shows how one could \textit{in principle} distil a small message that has fallen into a large black hole, which has evaporated past the Page time, after a very short time, the scrambling time $t_s\sim \beta \log S$, by collecting just a few more Hawking quanta than what the original message consisted of.
One of the requirements for a successful decoding is that one has to have access to all the early Hawking radiation emitted before the Page time.
Somewhat surprisingly, it follows that the early radiation purifies the small amount of late Hawking radiation and hence allows for information recovery. 
However, the authors stress that the actual decoding of the Hawking quanta could be very complicated.\footnote{Recently Yoshida and Kitaev gave a quantitative construction of the decoding procedure for the Hayden-Preksill protocol and evaluate its corresponding complexity in \cite{Yoshida:2017non}.}

It has recently turned out that, within the context of nearly $AdS_2$ gravity, there exists a precise formulation of the Hayden-Preskill thought experiment in which the decoding operation is actually relatively simple \cite{Maldacena:2017axo}.
This argument is based on the observation that a non-local and time-dependent coupling between the two copies of a large $N$ CFT in the thermofield double state renders the wormhole of the eternal black hole traversable \cite{Gao:2016bin}.\footnote{See \cite{Caceres:2018ehr} for the generalization to rotating BTZ black holes.}
In particular, the non-local coupling can be seen to generate averaged negative null energy in the bulk in the form of a negative energy shockwave \cite{Maldacena:2017axo}.
The gravitational backreaction of shockwaves is well-known to shift geodesics of particles that cross the shock \cite{Aichelburg:1970dh,Dray:1984ha}.
Due to the fact that the energy of the shockwave is negative, it allows the infalling message to cross regions in the eternal black hole.
This is illustrated in Figure \ref{fig:singlens}.
Since the message apparently falls into the black hole, the crossing between the two regions can be interpreted as the recovery of information from Hawking radiation in the original thought experiment \cite{Maldacena:2017axo}.
The decoding is simple in this set-up because the infalling message is not scrambled by the black hole, in contrast to the original assumption of Hayden and Preskill.
Instead, from the bulk perspective, the message essentially falls freely through empty space apart from only briefly encountering a shock of negative energy.
Therefore, upon including the non-local coupling in the CFT, this example presents a manifestly unitary bulk perspective on the recovery of information from a black hole.


In this work, we revisit this mechanism in the BTZ geometry.
The dual of the eternal BTZ black hole is well known to be the thermofield double state in a CFT$_2$ \cite{Maldacena:2001kr}.
We consider the infinite boost limit of a specific double trace operator which couples the left and right CFT.
This non-local coupling is a slight modification of the one considered in \cite{Gao:2016bin}.
In the infinite boost limit, we will show that the stress-energy tensor in the background of the non-local coupling becomes that of a negative energy shockwave, moving from region I into region II.
A small energy message then, released early enough from the boundary in region II, will have a large relative boost with respect to the non-local coupling and will be highly blueshifted when it reaches the mouth of the wormhole.
Therefore, one has to take into account its gravitational backreaction.
We do this by treating the message as a positive energy shock wave \cite{Shenker:2013pqa}.
This motivates us to construct the BTZ geometry with both a positive and negative energy shockwave by making use of the patching construction of \cite{Dray:1985yt, Shenker:2013pqa}.
We show that the backreaction of the message results in a reduction of the traversability of the wormhole, as also proposed in \cite{Maldacena:2017axo}.
This allows us to determine an upper bound on the amount of information carried by the message, which turns out to be only an $\mathcal{O}(1)$ amount of (qu)bits.
Even though this is perhaps disappointing, we argue that it is consistent with the late-time results of \cite{Maldacena:2017axo} and also suggest avenues for improving this bound.

Analysis of our geometry shows in addition that the combined backreaction of the negative and positive energy shockwaves increase the BTZ mass on one side of the Kruskal diagram, whereas it is decreased on the other side.
Interpreting the extended BTZ spacetime as a geometrization of the pure state of a black hole and its radiation at the Page time \cite{Maldacena:2013xja}, we propose an interpretation of the decrease of the mass on one side in terms of the second half of the Page curve.
We relate the simultaneous increase on the other side to a rejuvenation of the black hole before a scrambling time, as recently discussed in a different context in \cite{Almheiri:2019psf,Penington:2019npb}.

Motivated by this basic feature of the two-shock geometry, we construct a multiple shockwave geometry, which provides a set-up with which we are able to reflect the second half of the Page curve more accurately.
We estimate the amount of information in each shock and argue that the resulting curve, which measures the time-dependence of traversability in this geometry, is closely related to the Page curve of an evaporating black hole after the Page time. 
In the end, we find that our simple geometrical model allows us to reproduce some of the results derived using quantum field theoretic methods in \cite{Maldacena:2017axo,Almheiri:2019psf,Penington:2019npb}.

Finally, we employ our backreacted geometry to revisit the computation of the scrambling time by using the mutual information method of \cite{Shenker:2013pqa}. 
We will see that the negative energy shockwave results in a delay of the scrambling time, providing quantitative support for the idea that the traversable wormhole prevents the information from fully randomizing (contrary to the assumption in the Hayden-Preskill protocol). 

The rest of this paper is organized as follows.
In Section \ref{sec:non-local}, we compute the (averaged) null energy created by adding a suitable non-local coupling in the bulk theory of a scalar field theory minimally coupled to Einstein gravity. 
In Section \ref{sec:patchmetric}, we construct the exact metric of the backreacted BTZ by two shock waves, using the method of \cite{Dray:1985yt}. 
This allows us in Sections \ref{sec:bound} and \ref{sec:bh-evap} to study the bound on information that can be sent through the wormhole and a toy model for black hole evaporation respectively. 
Finally, in Section \ref{sec:scrambling} we show how the negative energy shockwave delays the scrambling time. 
We will conclude in Section \ref{sec:disc} with some open questions.




 
\section{Non-local couplings and negative null energy}\label{sec:non-local}

We shall study Einstein gravity in 2+1 dimensions with  negative cosmological constant and its dual CFT. It is well-known that black hole solutions exist in this theory, the so-called BTZ black holes \cite{Banados:1992wn}.  Despite the relative simplicity of these black holes, they have many interesting features in common with their higher dimensional cousins. Here we focus on the neutral, eternal BTZ black hole. Its Kruskal extension has two asymptotically AdS$_3$ regions and is dual to the thermofield double (TFD) state in the two CFTs \cite{Maldacena:2001kr}.
Some of the necessary background on BTZ black holes is reviewed in Appendix \ref{sec:BTZmetricscoordinate}. 

The wormhole connecting the two asymptotically AdS$_3$ regions is non-traversable, which is the bulk manifestation of the fact that no information can be communicated between the two (decoupled) CFTs.
Another perspective is that the entanglement between the two CFTs is not enough to transmit information.
Instead, as for instance in a quantum teleportation protocol, one requires in addition a communication channel between the CFTs.
Applying this intuition, a relevant double trace deformation was considered in \cite{Gao:2016bin} that couples the two boundary CFTs:
\begin{equation}
\delta S = \int dtdx\, h(t,x)\mathcal{O}_R(t,x)\mathcal{O}_L(-t,x),
\end{equation}
where $\mathcal{O}_{R,L}$ are scalar operators in the right and left boundary theory of dimension less than $1$ and $h(t,x)$ is a time-dependent coupling, vanishing before $t=t_0$.
It was then shown that for a specific sign and to first order in the coupling, the averaged null energy becomes negative:
\begin{equation}
\int^{\infty}_{U_0}dU\, T_{UU}<0.
\end{equation}
This provides a sufficient condition for the wormhole to become traversable.

\begin{figure}[t]
	\centering
	\includegraphics[ trim=1.5cm 14.5cm 7cm 3cm, width=0.5\textwidth]{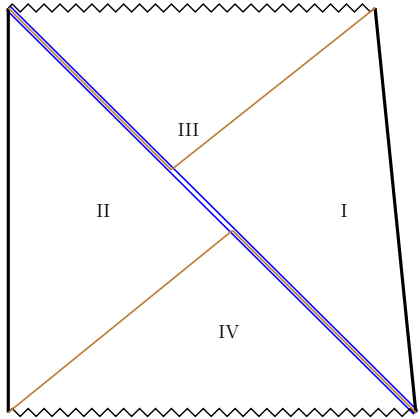}
	\caption{\small \it The Penrose diagram showing the gravitational backreaction of a single negative energy shock wave (blue double line) on the BTZ spacetime. This yields a traversable passage from region II to region I.}
	\label{fig:singlens}
\end{figure}

A more concrete and for us important interpretation of the negative energy was given in \cite{Maldacena:2017axo}.
They showed that a similar coupling in the context of nearly $AdS_2$ gravity can be thought of as releasing two negative energy shockwaves into the bulk. 
The gravitational backreaction of such shockwaves is well-known to shift the geodesics of crossing particles \cite{Aichelburg:1970dh,Dray:1984ha}.
From this perspective, it is particularly clear that a probe particle sent in early enough from either boundary is able to reach the other boundary due to the shift induced by the shockwave.
As emphasized in \cite{Maldacena:2017axo}, this holds for any bulk field, not just the one dual to $\mathcal{O}$, since the effect is mediated through gravitational backreaction.
See Figure \ref{fig:singlens}.

\begin{figure}[t]
	\centering
	\includegraphics[ trim=1cm 13cm 4cm 2cm, width=0.6\textwidth]{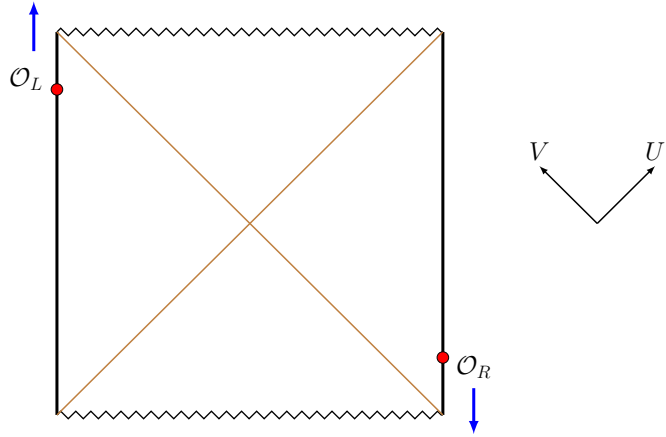}
	\caption{\small \it Infinite boost limit of non-local coupling ${\cal O}_L(0, \phi_0){\cal O}_R(0, \phi_0')$.}
	\label{fig:btz-boost}
\end{figure}

In order to make the interpretation of the non-local coupling in terms of negative energy shockwaves transparent in the context of the BTZ black hole, we first place the operators $\mathcal{O}_{L,R}$ at $t=0$ on their respective boundaries.
We then move to an infinitely boosted frame in which $\mathcal{O}_{L,R}$ are located at late left and right time respectively.
This is illustrated in Figure \ref{fig:btz-boost}.
In this case, we expect the coupling to generate a single shockwave, with respect to a fixed frame, that propagates along the $U$-axis.

More precisely, the coupling between the boundary CFTs we propose is given by:\footnote{There is a subtle order of limits between the infinite boost limit and the $\epsilon\to 0$ limit, the latter appearing during the computation of the expectation value of the stress-energy tensor. We have in mind that $\epsilon\to 0$ first, as will be discussed in Section \ref{ssec:comp-em-tensor} in more detail.}
\begin{equation}\label{eq:uninegrated-coupling}
\delta S =i\alpha \lim_{\tau_0\to-\infty}U_0\int_0^{2\pi} d\phi_0\int_0^{2\pi} d\phi_0'
\, h(\phi_0, \phi_0'){\cal O}_L(\tau_0, \phi_0){\cal O}_R(\tau_0, \phi_0').
\end{equation}
Besides the infinite boost, as reflected in the limit of $\tau_0\to- \infty$, this proposal differs from \cite{Gao:2016bin} in four respects: the two operators can be located at different transverse positions in $\phi$, there is no sign flip in the time coordinate for the left operator, the coupling is localized in time and finally there is an explicit appearance of the Kruskal coordinate $V_0$. 

We will be interested in the computation of the expectation value of the energy-momentum tensor of a massive scalar field in the bulk in the presence of \eqref{eq:uninegrated-coupling}.
The general philosophy of our computation closely follows \cite{Gao:2016bin} but the details differ.

The general form of the energy-momentum tensor of a massive scalar field reads:
\begin{equation}
T_{\mu\nu}=\partial_{\mu}\phi\partial_{\nu}\phi-\frac{1}{2}g_{\mu\nu}\partial_{\rho}\phi\partial^{\rho}\phi-\frac{1}{2}g_{\mu\nu}m^2\phi^2 .
\end{equation}
Its expectation value in the deformed BTZ spacetime can be expressed in terms of the associated two-point function $G(x,x')$ as:
\begin{equation}\label{eq:stress-tensor-ev}
\langle T_{\mu\nu}\rangle=\lim_{x\to x'}\biggl[\partial_{\mu}\partial'_{\nu}G(x,x')-{1\over 2}g_{\mu\nu}\partial_{\rho}\partial'^{\rho}G(x,x')-\frac{1}{2}g_{\mu\nu} m^2G(x,x')\biggr].
\end{equation}
We thus wish to compute 
\begin{align}
G_L(x,x')=\langle\phi_L(x)\phi_L(x')\rangle_{\rm TFD, \rm def}\qquad\quad\mbox{and}\qquad\quad G_R(x,x')=\langle\phi_R(x)\phi_R(x')\rangle_{\rm TFD, \rm def}\ .
\end{align}
In fact, both choices lead to the same result because the only difference between the operators in the left and right wedges is the $i\beta/2$ shift in time under which $(U, V)$ maps to $(-U, -V)$ (see \eqref{eq:coordinates} in Appendix \ref{sec:BTZmetricscoordinate}). 
Following \cite{Gao:2016bin}, we will adopt the interaction picture to deal with the deformation:
\begin{align}
G(x,x')=\langle\phi^{\rm H}_R(x)\phi^{\rm H}_R(x')\rangle_{\rm TFD, \rm def}=\langle U(\tau)^{-1}\phi^{\rm I}_R(x)U(\tau)U(\tau')^{-1}\phi^{\rm I}_R(x')U(\tau')\rangle_{\rm TFD}
\end{align}
with $U\equiv {\cal T}\exp\left(-i\int_{-\infty}^{\tau}d\tau_1 \delta H_{\rm shock}(\tau_1)\right)$ the time evolution operator and the deformation Hamiltonian given by: 
\begin{equation}
\delta H_{\rm shock}=-\alpha\!\lim_{\tau_0\to -\infty}\!U_0\,\delta(\tau_1-\tau_0)\int_0^{2\pi} d\phi_0\int_0^{2\pi} d\phi_0'
\, h(\phi_0, \phi_0'){\cal O}_L(\tau_0, \phi_0){\cal O}_R(\tau_0, \phi_0').
\end{equation}
To first order in $\alpha$ we have
\begin{align}
\begin{split}
G_{\alpha}(x,x')=&-i\alpha \Biggl[ U_0 \langle  \left[{\cal O}_L(\tau_0, \phi_0){\cal O}_R(\tau_0, \phi_0'), \phi_R(x)\right]\phi_R(x') \rangle_{\rm TFD}\\
&\hspace{1.5cm}+U'_0\left\langle \phi_R(x)\left[{\cal O}_L(\tau_0, \phi_0){\cal O}_R(\tau_0, \phi_0'), \phi_R(x')\right]\right\rangle_{\rm TFD}\Biggr]\\
=&-i\alpha\Biggl[U_0\left\langle {\cal O}_L(\tau_0, \phi_0)\left[{\cal O}_R(\tau_0, \phi_0'), \phi_R(x)\right]\phi_R(x')\right\rangle_{\rm TFD}
+(\tau\leftrightarrow\tau')\Biggr]
\end{split}
\end{align}
where we omitted the superscript I, the limit of $\tau_0$, $h(\phi_0,\phi_0')$ and the integrals over $\phi_0$ and $\phi_0'$. 
The points $x$ and $x'$ are $x=(\tau, r,\phi)$ and $x'=(\tau', r', \phi')$.
In the second line, we used causality $[\phi_L, \phi_R]=0$ for any locations of $\phi_L$ and $\phi_R$ in the extended BTZ spacetime. 
Finally, for the free field (in the absence of gravitational mediation), this four-point function factorizes into:\footnote{In the CFT, this is equivalent to the large $N$ factorization.}
\begin{equation}\label{eq:2ptfunction-btz-def}
G_{\alpha}(x,x')\simeq -i\alpha\Biggl[U_0\left\langle {\cal O}_L(\tau_0, \phi_0)\phi_R(x')\right\rangle_{\rm TFD}\left\langle\left[{\cal O}_R(\tau_0, \phi_0'), \phi_R(x)\right]\right\rangle_{\rm TFD}
+(\tau\leftrightarrow\tau')\Biggr] .
\end{equation}
Notice that the discussion so-far has been very general, and in particular applies to general dimensions.
In the next subsections we are going to compute the stress-energy tensor in the BTZ spacetime.

\subsection{Wightman functions in BTZ}
\label{sec:Wightman}

As can be seen from \eqref{eq:2ptfunction-btz-def}, we will only need the bulk-to-boundary Wightman functions. 
They can be constructed from the bulk-to-boundary correlation functions in AdS$_3$, which are given in terms of AdS-Schwarzchild coordinates as \cite{Ichinose:1994rg,KeskiVakkuri:1998nw,Gao:2016bin}:\footnote{We shall set the AdS radius $\ell=1$ in this section unless stated otherwise.}
\begin{align}
\hspace{-.8cm}
K_{\Delta}(\tau, r,\phi ; \tau_0,\phi_0)={R^{\Delta}\over 2^{\Delta+1}\pi}\left[-{\sqrt{r^2-R^2}\over R}\cosh\left(R(\tau-\tau_0)\right)
+{r\over R}\cosh(R(\phi-\phi_0))\right]^{-\Delta}\!\!\!\!\!\!\! .
\end{align}
By replacing $\phi\to \phi+2\pi n$ and summing over images labeled by $n$ one obtains the Wightman functions for the BTZ spacetime:
\begin{equation}
K^{\rm BTZ}_{\Delta}(\tau, r,\phi ; \tau_0,\phi_0)=\sum_{n=-\infty}^{\infty}K_{\Delta}(\tau, r,\phi +2\pi n; \tau_0,\phi_0).
\end{equation}
For the computation of the stress-tensor, this amounts to extending the integration range of $\phi$ from $[0, 2\pi]$ to $(-\infty, +\infty)$.
Notice that we have:
\begin{equation}\label{eq:identityBTZ}
-{\sqrt{r^2-R^2}\over R}\cosh\left(R(\tau-\tau_0)\right)={2(UV_0+VU_0)\over(1+UV)(1-U_0V_0)}\ ,
\end{equation}
where $U_0V_0=-1$ and conventions may be found in Appendix \ref{sec:BTZmetricscoordinate}.
This allows us to express the correlation function as follows:
\begin{align}\label{KDeltaboost}
\begin{split}
K_{\Delta}(\tau, r,\phi ; \tau_0,\phi_0)
&= {R^{\Delta}\over 2^{\Delta+1}\pi}
{(1+UV)^{\Delta}\over\left[-(U-U_0)(V-V_0)+2(1-UV)\sinh^2\left({R(\phi-\phi_0)\over 2}\right)\right]^{\Delta}}\ .
\end{split}
\end{align}
In our computation, we will focus on the right wedge, meaning $V<0$ and $U>0$.
For the Wightman functions, $U-U_0$ and $V-V_0$ are shifted by $+i\epsilon/2$ or $-i\epsilon/2$, depending on the specific ordering of operators.

\subsection{Computation of stress-energy tensor}\label{ssec:comp-em-tensor}

At this point, we are ready for the computation of the expectation value of $T_{UU}$.
Since for the BTZ spacetime $g_{UU}=0$, the formula \eqref{eq:stress-tensor-ev} simplifies to:
\begin{equation}\label{eq:stress-tensor-vv-cpt-ev}
\langle T_{UU}\rangle=\lim_{x\to x'}\partial_{U}\partial_{U'}G(x,x').
\end{equation}
It will be convenient to express the bulk-to-boundary correlation function in an integral representation:
\begin{align}
\hspace{-.2cm}K_{\Delta}(\tau, r,\phi ; \tau_0,\phi_0)= {R^{\Delta}(1+UV)^{\Delta}\over 2^{\Delta+1}\pi\Gamma(\Delta)}\int_0^{\infty}\!\!ds
s^{\Delta-1}e^{-s\left[-(U-U_0)(V-V_0)+2(1-UV)\sinh^2\!\!{R(\phi-\phi_0)\over 2}\right]}.\!\!
\end{align}
To compute \eqref{eq:2ptfunction-btz-def}, we first note that the two-point functions are given by:
\begin{align}
\hspace{-.3cm}\left\langle {\cal O}_L(\tau_0, \phi_0)\phi_R(x')\right\rangle_{\rm TFD}&={R^{\Delta}(1+U'V')^{\Delta}\over 2^{\Delta+1}\pi\Gamma(\Delta)}\int_0^{\infty}\!\!ds
s^{\Delta-1}e^{-s\left[-(U'+U_0)(V'+V_0)+2(1-U'V')\sinh^2\!\!{R(\phi'-\phi_0)\over 2}\right]}\ ,\nonumber\\
\hspace{-.7cm}
\left\langle[{\cal O}_R(\tau_0, \phi'_0), \phi_R(x)]\right\rangle_{\rm TFD}&={R^{\Delta}(1+UV)^{\Delta}\over 2^{\Delta+1}\pi\Gamma(\Delta)}\int_0^{\infty}\!\!ds
s^{\Delta-1}\biggl[e^{-s\left[-\left(U-U_0-{i\epsilon\over 2}\right)\left(V-V_0-{i\epsilon\over 2}\right)+2(1-UV)\sinh^2\!\!{R(\phi-\phi'_0)\over 2}\right]}\nonumber\\
&\hspace{2.2cm}-e^{-s\left[-\left(U-U_0+{i\epsilon\over 2}\right)\left(V-V_0+{i\epsilon\over 2}\right)+2(1-UV)\sinh^2\!\!{R(\phi-\phi'_0)\over 2}\right]}\biggr].
\label{eq:2pt-functions-btz-explicit}
\end{align}
Since $U_0$ and $V_0$ refer to boundary coordinates, using the equation \eqref{eq:UVtauboundary} from Appendix \ref{sec:BTZmetricscoordinate}, they are simply the boost factors $V_0= -e^{-R\tau_0}$ and $U_0= e^{R\tau_0}$.
Hence, in the $\tau_0\to-\infty$ limit, we have $U_0\to 0^+$ and $V_0\to -\infty$ while $U_0V_0=-1$.

Anticipating the infinite boost limit, we now plug in \eqref{eq:2pt-functions-btz-explicit} into \eqref{eq:stress-tensor-vv-cpt-ev} while keeping only the dominant contribution in $V_0$:
\begin{align}\label{eq:vv-derivs-two-point-function-1}
\begin{split}
I_\Delta=&\,\,U_0\int_0^{\infty}\!\!ds_1
s_1^{\Delta}(V'+V_0)e^{-s_1\left[-(U'+U_0)(V'+V_0)+2(1-U'V')\sinh^2\!\!{R(\phi'-\phi_0)\over 2}\right]}\\
&\times \int_0^{\infty}\!\!ds_2
s_2^{\Delta}\Biggl[\left(V-V_0-{i\epsilon\over 2}\right)e^{-s_2\left[-\left(U-U_0-{i\epsilon\over 2}\right)\left(V-V_0-{i\epsilon\over 2}\right)+2(1-UV)\sinh^2\!\!{R(\phi-\phi'_0)\over 2}\right]}\\
&\hspace{2.2cm}
-\left(V-V_0+{i\epsilon\over 2}\right)e^{-s_2\left[-\left(U-U_0+{i\epsilon\over 2}\right)\left(V-V_0+{i\epsilon\over 2}\right)+2(1-UV)\sinh^2\!\!{R(\phi-\phi'_0)\over 2}\right]}\Biggr]
\end{split}
\end{align}
where we dropped some of the prefactors which we will reinstate at the end. $I_{\Delta}$ is thus equal to $\partial_{U}\partial_{U'}G(x,x')$ up to some factors in the $\epsilon\to 0$ and $\tau_0\to-\infty$ limits. 
From this expression, it can be seen why we included a factor of $U_0$ in the non-local coupling \eqref{eq:uninegrated-coupling}.
It precisely compensates the diverging factor $(V'+V_0)$, in the limit, to yield a finite answer, as we will see below in more detail.

To see how the shockwave stress tensor emerges from this expression, we first perform a change of the integration variable:
\be
s_2=\left(-\left(U-U_0\mp{i\epsilon\over 2}\right)\left(V-V_0\mp{i\epsilon\over 2}\right)\right)^{-1}u_2\ .
\ee
The resulting expression reads:
\begin{align}\label{eq:vv-derivs-two-point-function-2}
\begin{split}
\hspace{-.2cm}I_\Delta=&\,\,U_0\int_0^{\infty}\!\!ds_1
s_1^{\Delta}(V'+V_0)e^{-s_1\left[-(U'+U_0)(V'+V_0)+2(1-U'V')\sinh^2\!\!{R(\phi'-\phi_0)\over 2}\right]}\\
&\times \int_0^{\infty}\!\!du_2
u_2^{\Delta}\Biggl[{\left(-\left(U-U_0-{i\epsilon\over 2}\right)\left(V-V_0-{i\epsilon\over 2}\right)\right)^{-\Delta}\over U-U_0-{i\epsilon\over 2}}
e^{-u_2\left[1+{2(1-UV)\sinh^2\!\!{R(\phi-\phi'_0)\over 2}\over -\left(U-U_0-{i\epsilon\over 2}\right)\left(V-V_0-{i\epsilon\over 2}\right)}\right]}\\
&\hspace{1.5cm}
-{\left(-\left(U-U_0+{i\epsilon\over 2}\right)\left(V-V_0+{i\epsilon\over 2}\right)\right)^{-\Delta}\over U-U_0+{i\epsilon\over 2}}
e^{-u_2\left[1+{2(1-UV)\sinh^2\!\!{R(\phi-\phi'_0)\over 2}\over -\left(U-U_0+{i\epsilon\over 2}\right)\left(V-V_0+{i\epsilon\over 2}\right)}\right]}\Biggr]
\end{split}
\end{align}
As already alluded to above, before taking the infinite boost limit of this expression, we first take $\epsilon\to 0$.
An identity that will help to proceed is given by:
\begin{equation}\label{eq:cauchy-relation}
\lim_{\epsilon\to 0}\frac{1}{x\pm i\epsilon}=\mathcal{P}\left(\frac{1}{x}\right)\mp i\pi \delta(x),
\end{equation}
where $\mathcal{P}$ denotes the Cauchy principal value.
Since we will ultimately take the infinite boost limit, in which in particular $V_0\to- \infty$, whereas we keep $V$ fixed, it is clear that we can safely take the $\epsilon\to 0$ limit in the terms $(V-V_0\pm \frac{i\epsilon}{2})$.
On the other hand, we have to be more careful with the terms $(U-U_0\pm \frac{i\epsilon}{2})$, since $U-U_0$ may well be vanishing.
It turns out that one can essentially replace the $(U-U_0 \pm \frac{i\epsilon}{2})$ in the denominator of \eqref{eq:vv-derivs-two-point-function-2} by \eqref{eq:cauchy-relation} while taking $\epsilon\to 0$ in the other terms.\footnote{In fact, we can show that for positive integers $n$
\begin{equation}
\lim_{\epsilon\to 0}{\frac{1}{(x+i\epsilon)^n}} = \mathcal{P}  \left(\frac{1}{x^n} \right) + C_{n} \frac{\delta(x)}{x^{n-1}}
= \mathcal{P}  \left(\frac{1}{x^n} \right) + C_{n}{(-1)^{n-1}\over (n-1)!}\delta^{(n-1)}(x)
	\end{equation} for some constant $C_n \in \mathbb{C}$ by multiplying the factor $(x-i\epsilon)^n$ in both the numerator and denominator and expanding it in the numerator and making use of the fact \begin{equation}\label{eq:epsilondelta}
\lim_{\epsilon\to 0} \frac{\epsilon^m}{(x^2+\epsilon^2)^n} = \frac{\delta(x)}{x^{2n-m-1}} \frac{\Gamma(\frac{m}{2})\Gamma(\frac{1}{2}) \Gamma(2n-m-1)}{2^{2n-m-1} \Gamma(n-\frac{m+1}{2})\Gamma(n)}\ .
	\end{equation}
	Note that each term in the expansion of $(x-i\epsilon)^n$ results in the same distribution $x^{1-n}\delta(x)$. 
	More generally, it can be shown that
	\be
	{\rm Im}(x+i\epsilon)^{-\nu}={\rm Im}{(x-i\epsilon)^{\nu}\over (x^2+\epsilon^2)^{\nu}}\propto x^{1/2+n-\Delta_n}\delta^{(n)}(x)\ ,
	\ee
	for $\nu=\Delta_n+1/2$ with non-negative integers $n$ and $1/2+n\le\Delta_n\le 3/2+n$.
	Thus except for $\Delta= 1/2$, the prescription used here determines \eqref{eq:vv-derivs-two-point-function-3} only up to a constant which, however, is not relevant in our application. 
}
In this case, it is not difficult to see that an overall $\delta(U-U_0)$ factor appears in \eqref{eq:vv-derivs-two-point-function-2}.
We use this fact to simplify the term in the exponent:
\begin{align}
\begin{split}
e^{-u_2\left({2(1-UV)\sinh^2\!\!{R(\phi-\phi'_0)\over 2}\over -\left(U-U_0\right)\left(V-V_0\right)}\right)}\stackrel{U-U_0\to 0}{\simeq} \frac{2}{R}\sqrt{{-\left(U-U_0\right)\left(V-V_0\right)\over 2\pi u_2(1-UV)}} \delta(\phi-\phi'_0)
\end{split}
\end{align}
where we used:
\begin{equation}
e^{-{\sinh^2x\over\eta}}\stackrel{\eta\to 0^{+}}{\simeq}e^{-{x^2\over\eta}}\simeq \sqrt{\eta\over\pi}\delta(x)\ .
\end{equation}
This finally leaves us with the following object:
\begin{align}\label{eq:vv-derivs-two-point-function-3}
\begin{split}
I_\Delta\simeq&\,\,{4i\pi U_0\delta(U-U_0)\over R\sqrt{2\pi(1-UV)}}\int_0^{\infty}\!\!ds_1
s_1^{\Delta}(V'+V_0)e^{-s_1\left[-(U'+U_0)(V'+V_0)+2(1-U'V')\sinh^2\!\!{R(\phi'-\phi_0)\over 2}\right]}\\
&\hspace{.5cm}\times \delta(\phi-\phi'_0)\int_0^{\infty}\!\!du_2
u_2^{\Delta-1/2}e^{-u_2}\left(-\left(U-U_0\right)\left(V-V_0\right)\right)^{\half-\Delta}.
\end{split}
\end{align}
This expression clearly only makes sense for $\Delta=\frac{1}{2}$: it vanishes for $\Delta<\frac{1}{2}$ and diverges for $\Delta>\frac{1}{2}$.\footnote{To be more precise, as implied in the previous footnote, it is singular as $(U-U_0)^{1/2+n-\Delta_n}\delta^{(n)}(U-U_0)$ with $1/2+n \le\Delta_n\le 3/2+n$. However, at the special values of conformal dimension $\Delta_n=1/2+n$, the factor $(U-U_0)^{1/2+n-\Delta_n}=1$ and it is not as singular and becomes simply the $n$-th derivative of the $\delta$-function.}
For $\Delta=\frac{1}{2}$, we are now ready to take the infinite boost limit.
Since we have in the limit that $U_0V_0=-1$ and keeping in mind that $U'\to U$, $V'\to V$ and $\phi'\to\phi$, one finds:
\begin{align}
\begin{split}
I_{\Delta=1/2}&\stackrel{\tau_0\to- \infty}{\longrightarrow}\,\,-{4i\pi\delta(U)\over R\sqrt{2\pi(1-UV)}}\delta(\phi-\phi'_0)\int_0^{\infty}\!\!ds_1
s_1^{\frac{1}{2}}e^{-2s_1\left[1+(1-U'V')\sinh^2\!\!{R(\phi'-\phi_0)\over 2}\right]}\ .
\end{split}
\end{align} 
Reinstating the prefactors we dropped in $I_{\Delta}$, we finally obtain that
\begin{align}\label{eq:TVVgen}
\begin{split}
\langle T_{UU}\rangle ={\alpha\delta(U)\over R\sqrt{2}\pi^{5/2}}\int_{-\infty}^{\infty} d\phi_0
\, &h(\phi_0, \phi)\int_0^{\infty}\!\!ds_1
s_1^{\frac{1}{2}}e^{-2s_1\cosh^2\!\!{R(\phi-\phi_0)\over 2}}\ .
\end{split}
\end{align}
For a spherical shell, the case we are most interested in, the coupling function $h(\phi_0, \phi)$ must be a constant: for a constant $h$, the $\phi$-dependence can be eliminated by a shift of $\phi_0$ and the stress-tensor is independent of $\phi$. After carrying out the $s_1$ and $\phi_0$ integrals, by choosing the constant $h=4R/G_N$, we can bring \eqref{eq:TVVgen} into the canonically normalized form
\be
\langle T_{UU}\rangle ={\alpha\over 4\pi G_N}\delta(U)\ .
\ee
The choice of a negative $\alpha$ yields a negative shockwave and thereby a traversable wormhole.

We should note that the fact that $\Delta=\frac{1}{2}$ is special was already noted in \cite{Gao:2016bin}.
In their computation, it is the value where the stress-tensor crosses over from being regular to being singular.
However, in their case the singularity is integrable and is therefore not very worrisome.
Since we have considered both the infinite boost limit and the limit in which the time interval collapses to a point (and is brought to $t=0$), we believe that the singular behaviour observed in \eqref{eq:vv-derivs-two-point-function-3} is more pronounced in the $\Delta>\frac{1}{2}$ case, whereas the value of the stress-energy for $\Delta<\frac{1}{2}$ is simply vanishing. 
We will further comment on this in Section \ref{sec:disc}.

\section{Traversable BTZ wormholes}  \label{sec:BTZWH}

In this section, we turn to the analysis on the gravity side. 
As we wish to study the global structure of the BTZ black hole, including an infalling observer passing through the horizon, we are going to work with the metric in the Kruskal coordinates introduced in Appendix \ref{sec:BTZmetricscoordinate}:  
\begin{equation}
ds^2 = \frac{-4\ell^2 dUdV +R^2(1-UV)^2 d\phi^2}{(1+UV)^2}
\end{equation}
where $\ell$ is the $AdS_3$ radius and $R$ is the horizon radius. 
The two boundaries are at $UV=-1$.
The event horizons are located at  $U=0$ and $V=0$ and there are (coordinate) singularities at $UV=1$. The black hole mass and entropy are given by $M =R^2/(8G_N\ell^2)$ and $S=\pi R/(2G_N)$, respectively, and its inverse Hawking temperature is $\beta_H=2\pi\ell^2/R$. 

\subsection{Patchwise construction of geometry} \label{sec:patchmetric}

To begin with, the single shock wave geometry in BTZ black holes is well-known and the metric is given by \cite{Hotta:1992qY, Shenker:2013pqa}
\be\label{SingleShock}
ds^2={-4\ell^2dUdV+R^2\left[1-U(V+\alpha\theta(U))\right]^2d\phi^2\over \left[1+U(V+\alpha\theta(U))\right]^2}
\ee
with the stress-energy tensor localized along the shock at $V=0$,
\be\label{TVValpha}
T_{UU}={\alpha\over 4\pi G_N}\delta(U)\ .
\ee
The positive energy shockwave (P-shock) corresponds to $\alpha>0$ whereas the negative energy shockwave (N-shock) corresponds to $\alpha<0$. The most important property of this spacetime is that the horizon $UV=0$ in the $U<0$ region is shifted to $U(V+\alpha\theta(U))=0$ in the $U>0$ region. 
This implies that an N-shock opens up a causal passage between the left and right Rindler wedges, rendering the wormhole traversable.
See Figure \ref{fig:singlens}.

Although we will not make use of it in this paper, it is sometimes useful to consider {\it discontinuous} coordinates, defined by $U=u$ and $V=v-\alpha\theta(u)$, in terms of which the metric takes the form  
\be\label{discontinuous}
ds^2={-4\ell^2dudv+4\ell^2\alpha\delta(u)du^2+R^2(1-uv)^2d\phi^2\over (1+uv)^2}\ .
\ee
In these coordinates the shift of the horizon is not manifest.
Instead the geodesics get shifted across the shock at $v=0$. In particular, the null geodesics are given by $v(u)=v_0+\alpha\theta(u)$ which jump across the shock by the amount of the shock wave strength $\alpha$. This is the coordinate system adopted in \cite{Maldacena:2017axo} and it is useful to keep this difference in mind when one relates our findings to theirs.

In Section \ref{sec:non-local}, we have shown that the infinite boost limit of certain double trace deformations which couple the left and right CFTs generate the stress-tensor of the $\delta$-function form \eqref{TVValpha}. 
Thus the N-shock geometry \eqref{SingleShock} with $\alpha=-|\alpha|$ is dual to the infinite boost limit of the nonlocal CFT coupling \eqref{eq:uninegrated-coupling} with an appropriate sign choice.

\begin{figure}[t]
\centering
	\includegraphics[trim=3cm 14cm 7cm 3.5cm,width=0.5\textwidth]{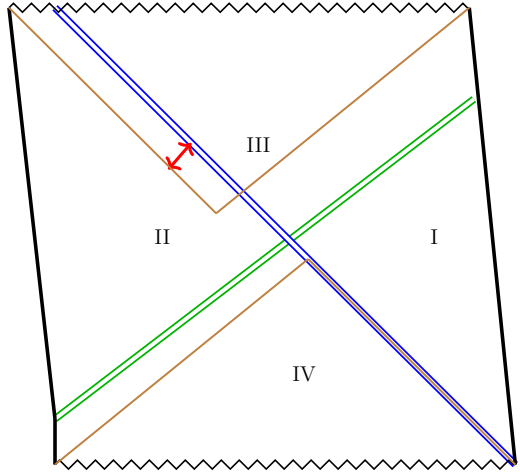}
	\caption{\textit{The non-local coupling induces a negative energy shock wave (blue double line) on the extended BTZ black hole background. The right future event horizon (brown solid line) is shifted forward to the future in Region III and as a consequence, the traversable passage opens and Region I and II become causally connected. The positive shock wave (green double line) as a message carrier is shot through the wormhole. Due to its backreaction, the left future horizon (brown solid line) is shifted back to the past  in Region II. As a result, the negative shock falls into the black hole, thereby reducing the black hole mass in Region III and enabling the message carrier to extract energy from the black hole.}  
	}
	\label{fig:NP-shock-penrose}
\end{figure}

With this result at hand, we now wish to spell out the mechanism of information transfer via traversable wormholes in an explicit and illustrative manner.
Due to the infinite boost, any message that is dropped into the black hole early enough will have a large relative boost with respect to the non-local coupling.
Consequently, we should take into account the gravitational backreaction of the message, which we model via a positive energy shock wave (P-shock) \cite{Shenker:2013pqa}.
This corresponds to the spacetime of two colliding spherical shells in BTZ, smeared along the $\phi$-circle, with one being a P-shock and the other being an N-shock.
The corresponding geometry is shown in Figure \ref{fig:NP-shock-penrose}.
In order to find the fully backreacted geometry of the two colliding shocks, we are going to use the patchwise construction developed by Dray and 't Hooft \cite{Dray:1984ha,Dray:1985yt}.

We focus on the two-shock geometry where the P-shock is at $V=a$ and the N-shock at $U=0$. 
The spacetime is divided into four regions separated by the N- and P-shocks as in Figure \ref{fig:NP-shock-penrose}:\footnote{We will use Arabic numbering for the regions of spacetime separated by shockwaves, while using the conventional Roman numbering for the causal regions.}
\begin{align}\label{regions}
{\rm Region\, \,1}:&\qquad U\ge 0\ ,\qquad V\le a \\
{\rm Region\,\,  2}:&\qquad U\le 0\ ,\qquad V\ge a \\
{\rm Region\,\,  3}:&\qquad U\ge 0\ ,\qquad V\ge a \\
{\rm Region\,\,  4}:&\qquad U\le 0\ ,\qquad V\le a\ .
\end{align}
The geometry in each region is locally the BTZ black hole:
\be \label{eq:localBTZ}
ds^2={-4\ell^2dU_idV_i+R_i^2\left(1-U_iV_i\right)^2d\phi^2\over \left(1+U_iV_i\right)^2}\ ,
\ee
where $i=1,2,3,4$ labelling each region of the spacetime. The shock waves influence only the future of the spacetime. Thus, Region 4 is unaffected and it is convenient to denote $(U_4, V_4)=(U, V)$ and $R_4=R$.
We will relate the coordinates in the future regions always to the coordinates $(U, V)$ and the horizon radius $R$.
In our setup, the N-shock is shot along the horizon $U=0$. This, in particular, implies that the horizon radius $R_1$ of Region 1 remains unaffected by the shock waves and we thus have $R_1=R$. 

The metric \eqref{eq:localBTZ} must be continuous across the shock waves. This requires the patching conditions: 
\begin{align}
{R_i(1-U_iV_i)\over 1+U_iV_i}={R_j(1-U_jV_j)\over 1+U_jV_j}\qquad\mbox{and}\qquad
{dU_idV_i\over \left(1+U_iV_i\right)^2}={dU_jdV_j\over \left(1+U_jV_j\right)^2}
\end{align}
for adjacent regions $i$ and $j$. By finding and solving the patching conditions, we can construct the whole spacetime of the colliding shock waves.
\paragraph{(1) Continuity between 1 \& 4:}
The continuity between Regions 1 and 4 requires 
\begin{align}
{1-U_1V_1\over 1+U_1V_1}={1-UV\over 1+UV}\Biggr|_{U=0}\qquad\mbox{and}\qquad
{dU_1dV_1\over \left(1+U_1V_1\right)^2}={dUdV\over \left(1+UV\right)^2}\Biggr|_{U=0}
\end{align}
in the region $V\le a$ since $R_1=R_4=R$.  These conditions are the same as those of the single shock geometry along the horizon.  Then we can readily show that they are solved by $(U_1,V_1) = (U,V+\alpha)$ for some constant $\alpha$ identified with the strength of the N-shock: as the N-shock hits the horizon $V=0$, it gets shifted by $\alpha$.
\paragraph{(2) Continuity between 2 \& 4:}
This case is more nontrivial. Again the continuity between regions 2 and 4 requires 
\begin{align}\label{eq:continuity-cond-2-4}
{R_2(1-U_2V_2)\over 1+U_2V_2}={R(1-UV)\over 1+UV}\Biggr|_{V=a}\qquad\mbox{and}\qquad
{dU_2dV_2\over \left(1+U_2V_2\right)^2}={dUdV\over \left(1+UV\right)^2}\Biggr|_{V=a}
\end{align}
 in the region $U\le 0$ where $U_2$ is regarded as a function of $U$ and $V_2$ as a function of $V$. We anticipate that as the P-shock hits the horizon at $U=0$, it gets shifted. This implies $R_2\ne R$.
 Now the first condition reads:
 \be
 U_2(U)={R_2-R+(R_2+R)aU\over V_2(a)\left(R_2+R+(R_2-R)aU\right)}
 \ .\label{124firstNo3}
 \ee
Since the shift of the horizon at $U=0$ in Region 2 must be given by the strength $\beta$ of the P-shock, 
we identify
\be
U_2(0)= {R_2-R\over V_2(a)(R_2+R)}= \beta\ .\label{124beta}
 \ee
Combining \eqref{124firstNo3} and the second condition in \eqref{eq:continuity-cond-2-4}, it is easy to show that $V_2'(a)aR=V_2(a)R_2$. Since we can consistently choose $V_2(V) =V+V_2(a)-a$, this becomes $aR=V_2(a)R_2$. As we are interested in the case when the P-shock is shot through the traversable window $0\le V\le |\alpha|$, we require $a>0$ and then we must have $V_2(a)>0$ so that $R, R_2>0$. For the P-shock $\beta > 0$, \eqref{124beta} implies that $R_2 > R$ and therefore $V_0\equiv a-V_2(a)>0$.
This is of course as expected, since the P-shock, being shot off the horizon, adds energy to the black hole in region II.
 
\paragraph{(3) Continuities between 1 \& 3 and 2 \& 3:} The continuity between regions 1 and 3 is along $V_1=a+\alpha$ in the region $U_1= U\ge 0$. 
In this patching we regard $V_3$ as a function of $V_1=V+\alpha$ and $U_3$ as a function of $U_1=U$. We then denote $V_3(a+\alpha)=c$. Similarly to the previous case, the first condition reads 
\be
U_3(U)={R_3-R+(R_3+R)(a+\alpha) U\over c\left(R_3+R+(R_3-R)(a+\alpha) U\right)}\ .\label{13firstNo2}
\ee
Combining this with the second condition requires
\be
V_3'(a+\alpha)(a+\alpha) R = c R_3\ ,\label{13constraintsII}
\ee
where the derivative $V_3'=\partial_1V_3$ is with respect to $V_1$.

Similarly, for the continuity between 2 \& 3 along $U=0$, or equivalently $U_2=\beta$ in the region $V\ge a$,  the first condition reads
\be
V_3(V_2)={R_3-R_2+(R_3+R_2)\beta (V-V_0)\over \gamma\left(R_3+R_2+(R_3-R_2)\beta (V-V_0)\right)}\ ,\label{23firstNo3}
\ee
where we used $V_2=V-V_0$ and defined $\gamma=U_3(\beta)$ when $U_3$ is regarded as a function of $U_2$. 
Combining this with the second condition yields
\be
U_3'(\beta)\beta R_2=\gamma R_3\ ,\label{23constraintsII}
\ee
where the derivative $U_3'=\partial_2U_3$ is with respect to $U_2$.
\paragraph{(4) Consistency:}From \eqref{13constraintsII}, \eqref{23constraintsII}, \eqref{13firstNo2} and \eqref{23firstNo3} we have
\be
V_3'(a+\alpha)U_3'(\beta)(a+\alpha)\beta {RR_2\over R_3^2}=c\gamma={R_3-R\over R_3+R}
={R_3-R_2+\beta V_2(a)(R_3+R_2)\over R_3+R_2+\beta V_2(a)(R_3-R_2)}\ .\label{consistencyNo3}
\ee
Using the definition of $\beta$ \eqref{124beta} and $aR =V_2(a)R_2$, we can infer that 
\be\label{eq:regionIImassbyshock}
\beta V_2(a)={R_2-R\over R_2+R}={a\beta R\over R_2}\quad\Longrightarrow\quad R_2={R\over 2}\left[1+a\beta+\sqrt{(1+a\beta)^2+4a\beta}\right]\ ,
\ee
Calculating $U_3'(\beta)$ and $V_3'(a+\alpha)$ from \eqref{13firstNo2}  and \eqref{23firstNo3}, we find that \eqref{consistencyNo3} reduces to 
\be
R_3^2-R^2=(a-|\alpha|)\beta \frac{R(R_2+R)^2}{R_2}\ ,\label{BHmassredTraversable}
\ee
where we used that $\alpha=-|\alpha|<0$ for the N-shock. This is negative in the traversable window $0<a<|\alpha|$.

The formulas \eqref{eq:regionIImassbyshock} and \eqref{BHmassredTraversable} present two of our main results. 
These consistency conditions give the relation between the black hole masses in respectively Region 2 and Region 3 and the initial black hole mass $M_0=R^2/(8G_N\ell^2)$ in Region 4.
As can be seen from Figure \ref{fig:NP-shock-penrose}, both Region 2 and 3 contain (parts of) the asymptotic boundaries of region I and II.
We emphasize that observers in the overlap of Region 2 and II see a black hole of mass $M_2=R_2^2/(8G_N\ell^2)$ and observers in the overlap of Region 3 and I see a black hole of mass $M_3=R_3^2/(8G_N\ell^2)$.

It can be seen from \eqref{BHmassredTraversable} that when the P-shock is sent through the traversable window, the entropy black hole in Region 3 $S=\pi R_3/(2G_N)$ is reduced due to the NP shock interaction $-|\alpha|\beta$. 
We can thus conclude that as we send messages via the traversable wormhole, this black hole evaporates!
The key to this process is the interaction between the message carrier (P-shock) and the traversable gate opener (N-shock).\footnote{In an earlier work \cite{Bak:2018txn}, the authors considered a similar process in the JT-model in two-dimensional dilaton gravity. They showed that the entropy decreases at a rate that respects the first law of black hole thermodynamics.}
We will explore this aspect of the NP shock interaction further in Sections \ref{sec:bound} and \ref{sec:bh-evap}.

For convenience, we summarize the coordinates $(U,V)$ in the four regions:
 \begin{align}\label{eq:BTZpatchsolution}
&(U_1, V_1)=(U, V-|\alpha|)\nn\\ 
&(U_2, V_2)=\left({R_2\left(R_2-R+(R_2+R)aU\right)\over aR\left(R_2+R+(R_2-R)aU\right)}, V-V_0\right)\\
&(U_3, V_3)=\left({R_3-R+(R_3+R)(a-|\alpha|) U\over c\left(R_3+R+(R_3-R)(a-|\alpha|) U\right)}, {R_3-R_2+(R_3+R_2)\beta (V-V_0)\over \gamma\left(R_3+R_2+(R_3-R_2)\beta (V-V_0)\right)}\right)\nn\\
&(U_4, V_4)=(U, V)\nn
\end{align}
where $V_0=a-V_2(a)>0$.

\subsection{Bound on transferable qubits}\label{sec:bound}

As detailed in the previous section, the gravitational backreaction of a negative energy shock wave makes the BTZ wormhole traversable.
Since the N-shock originates from an infinite boost limit of the boundary coupling, we have argued that one should also take into account the gravitational backreaction of a message sent from the left boundary at any finite time.
We have treated the backreaction of the message as that of a positive energy shockwave.
These two shockwaves divide the spacetime into four distinct BTZ regions and the patching together of these regions results in the consistency condition \eqref{BHmassredTraversable}.
In particular, we observed that the condition implies a decrease of the BTZ mass and hence entropy in Region 3 and an increase in Region 2, which contain parts of the asymptotic regions of Region I and II respectively.
This reduction in entropy motivates us to introduce a slight generalization of our set-up as a toy model for black hole evaporation in the next section.
However, to warm up, we will in this section first use the backreaction of the message to derive a bound on the amount of information that can be sent through the wormhole, in the spirit of \cite{Maldacena:2017axo}.

Let us look again at the formula for mass reduction:
\begin{align}
\begin{split}
{R_3^2-R^2\over R^2}=-(|\alpha|-a)\beta{(R_2(a,\beta)+R)^2\over RR_2(a,\beta)} \ge -1,
\end{split}
\label{BHmassredTraversable2}
\end{align}
where the lower bound arises from the requirement that $R_3\geq 0$ and $R_2(a,\beta)$ is given by \eqref{eq:regionIImassbyshock}.
Furthermore, we take the total momentum $\beta$ of the positive energy shockwave to consist of $N$ minimal momentum wavepackets:
\begin{equation}\label{eq:N-def}
\beta=N\beta_{\rm each}.
\end{equation}
Notice that this simple linear relation is justified by the linearity of superposition of shockwaves.\footnote{We will see this more explicitly in Section \ref{sec:bh-evap}, where we consider a multiple P-shock geometry.}
As in \cite{Maldacena:2017axo}, we propose that $N$ provides a measure for the amount of information that can be sent through the wormhole.
In other words, $N$ measures the traversibility of the wormhole. 

An upper bound on $N$ can be derived if in addition to the bound of \eqref{BHmassredTraversable2} we also have a lower bound on $\beta_{\rm each}$.
For this we invoke the uncertainty principle, following \cite{Maldacena:2017axo}.
For a signal sent at $V=a$, the coordinate distance to the future horizon is $\Delta V=|\alpha|-a$.
The uncertainty principle now determines a lower bound on the variance of the null momentum $\Delta P^U\sim \beta_{\rm each}$:\footnote{The bound on $\beta_{\rm each}$ may improve considerably in geometries with multiple wormholes connecting the left and right region, as emphasized in the recent work \cite{Bao:2019rjy}.}
\begin{equation}\label{eq:uncertainty-relation}
\Delta V \Delta P^U =(|\alpha|-a)\beta_{\rm each}\ge C,
\end{equation}
with some $\mathcal{O}(1)$ constant $C$.
This bound simply reflects the fact that for a finite width $\Delta V$, the momentum of a wavepacket should be large enough to fit through the wormhole.

\begin{figure}[t]
	\centering \includegraphics[height=2in]{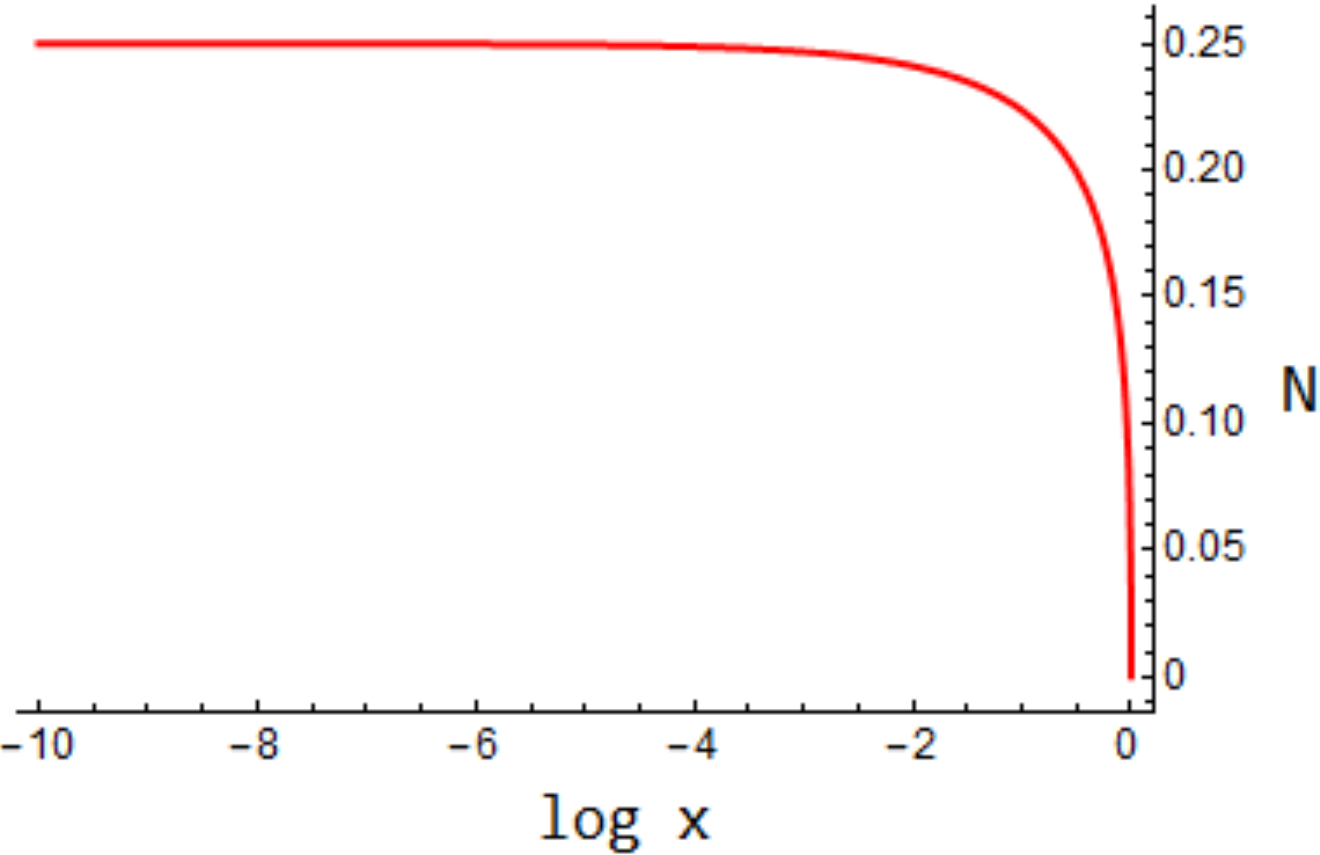} 
	\caption{\textit{The upper bound on the number of bits $N$ the message is allowed to contain is plotted for $C=1$. The horizontal axis is labeled by $\log x$ and parametrizes the AdS-Schwarzschild coordinate time when the message is sent. The earlier it is sent, the more bits it can carry.}}  \label{Nplot}
\end{figure}

Combining \eqref{BHmassredTraversable2} and \eqref{eq:uncertainty-relation}, we find:
\begin{equation}\label{eq:bound-on-bits}
N=\frac{\beta}{\beta_{\rm each}}\le \frac{1}{C}\left(\frac{2x-2+(2-x)\sqrt{1-x}}{x^2}\right),
\end{equation}
where $0\leq x\equiv a/|\alpha|\leq 1$.
This function decreases monotonically with $x$.
The two extrema are:
\begin{equation}\label{eq:Nbound-numbers}
{1\over 4}\,\,\ge\,\, N(x)C\,\,\ge\,\, 0.
\end{equation}
The upper bound is plotted in Figure \ref{Nplot} with respect to an AdS-Schwarzschild coordinate $t\equiv \log x$.
An important difference with the bound in \cite{Maldacena:2017axo} is that in our case the bound does not depend on the strength of the non-local coupling $\alpha$. 
We will comment on this in more detail below.

Before going into the actual value of the bound for $N$, let us first comment on the qualitative features of this plot.
To understand its form, we note that there are two main physical effects that contribute to the bound.
First of all, as can be seen from \eqref{BHmassredTraversable2}, the strength of the backreaction depends on $x$ and $\beta$ if we keep $\alpha$ fixed.
In particular, the backreaction is strong for small $x$ due to the fact that the earlier (in global time) the message is sent, the larger its relative boost is and hence its backreaction.
We will call this effect the \textit{boost effect}, which forbids large information transfer at very early times.
The strength of the backreaction also becomes stronger for large $\beta$ since this represents a larger energy of the message.
At this point, it may seem that messages sent at late times $x\sim 1$, having vanishingly small backreaction for arbitrary $\beta$, will lead to an enhancement of the amount of information $N$ that can be sent.
However, the reason this is not seen in the plot is due to the uncertainty relation \eqref{eq:uncertainty-relation}.
This effect, which we will call the \textit{uncertainty effect}, implies that when the message is sent at later times the energy of its component wavepackets $\beta_{\rm each}$ has to increase.
In fact, as can be seen from the plot, it turns out that the bound on $N$ decreases monotonically with $x$.
Concluding, one may view the curve in Figure \ref{Nplot} as resulting from two competing effects of which the uncertainty effect ``wins'' at all times.
However, the competition between the effects results in the fact that the curve is only slowly decreasing as a function of $x$.

The actual value for the bound shows that the number of bits that can be sent is $\mathcal{O}(1)$ for most values of $x$.
This may seem disappointing but is actually consistent with observations made in the context of nearly $AdS_2$ gravity in \cite{Maldacena:2017axo}.
In particular, due to the infinite boost of the non-local coupling our set-up corresponds for any $0\leq x\leq 1$ to a very late time case of theirs.
In such cases, the backreaction of the message for any finite time is strong.
Consequently, one expects only to be able to send a small amount of information through wormhole without destroying the fine correlations of the TFD state that sustain it.
Therefore, we see that our geometric analysis in terms of shockwave backreaction provides a simple framework to understand the late time results of the quantum field theoretic computations of \cite{Maldacena:2017axo}.

We expect that the bound can be improved upon by allowing the negative shock to propagate off the horizon, which would correspond to a smaller relative boost between the message and the non-local coupling.
In particular, this will lead to a dependence in the bound on $\alpha$, the strength of the non-local coupling, as also observed in \cite{Maldacena:2017axo}.
This set-up, however, will require us to revisit the computation of the stress tensor in Section \ref{sec:non-local}.
In particular, we would have to deal with a non-vanishing $T_{UU}$ component as well, originating form the boundary coupling in Region II.
We leave this analysis to future work.

An alternative perspective on the $\mathcal{O}(1)$ bound on $N$ has to do with the interpretation of the traversable wormhole as a simple gravitational perspective on the Hayden-Preskill protocol \cite{Maldacena:2017axo}.
Very briefly, the Hayden-Preskill protocol \cite{Hayden:2007cs} allows an observer, Bob, to retrieve information that has been thrown into an old black hole by Alice, after waiting a scrambling time, by collecting only a slightly larger amount of Hawking quanta than what the original message consisted of.
The requirements to perform the protocol are, apart from unitarity of black hole evaporation, that the black hole has evaporated past the Page time and that Bob has access to all of the early Hawking radiation, the exact black hole microstate and the exact dynamics.
However, even when these requirements are fulfilled, the precise decoding operation on the additional collected Hawking quanta is not specified and is presumably very complicated \cite{Yoshida:2017non}.

This protocol is realized in the traversable wormhole geometry of \cite{Maldacena:2017axo} as follows.
First of all, we take the point of view that the left side, i.e. Region II, of the extended black hole geometry represents the old black hole.
The early Hawking radiation, on the other hand, is represented by Region I. 
This representation makes use of the fact that the maximally entangled state of the black hole and early Hawking radiation at the Page time can be transformed into the thermofield double state, thus resulting in a double sided black hole \cite{Maldacena:2001kr,Maldacena:2013xja}.
Now, Alice's message is thrown into the black hole from the left boundary at some early time (which in our context is represented by the P-shock in Figure \ref{fig:NP-shock-penrose}).
Bob is thought to live on the right side, having access to the early radiation.
However, the non-local coupling allows him to collect late Hawking radiation from the left black hole.
This provides all the necessary requirements for a successful execution of the Hayden-Preskill protocol. 
The success of Bob's decoding in the present context is reflected by the passage of the message through the wormhole.
This decoding has not been very complicated; Bob only needs to set-up the non-local coupling and use a simple unitary to capture the message at the right boundary.
The bulk manifestation of this simple protocol is that the message passes through the wormhole essentially unaffected, apart from briefly encountering the N-shock.
A heuristic explanation for the efficacy of such a simple decoding protocol is that the traversable wormhole protocol delays the scrambling effect of the black hole and thereby evades the need for a complicated decoding operation.\footnote{We will show in Section \ref{sec:scrambling} by explicit computation that the negative shockwave indeed increases the scrambling time.}

Since in our set-up the message is sent at very early times, the delay in scrambling time by the N-shock is not strong enough to preclude a large scrambling effect.
This could provide an additional explanation of the $\mathcal{O}(1)$ bound on transferable qubits we have found.

Before closing this section, we should note that a message with $NC\lesssim \frac{1}{4}$, i.e. almost saturating the bound at early or intermediate times, has a very strong backreaction on the geometry.
In particular, from the formulas \eqref{eq:regionIImassbyshock} and \eqref{BHmassredTraversable} for $R_2$ and $R_3$, it can be easily checked that such a shock would result in $R_2\approx 1.05 R$ and $R_3\approx 0$.
This shows that our message increases the mass of the old black hole, proportional to $R^2_2$, whereas it almost completely reduces the ``size'' $R_3$ of the early radiation.
The reduction of $R_3$ can be interpreted as a reduction in the entanglement entropy of the early radiation subsystem, which we will denote by $S_R$.
This follows from the fact that, in the context of the thermofield double, the entanglement entropy of the reduced density matrix for a single boundary yields the area of the horizon via the Ryu-Takayanagi proposal \cite{Ryu:2006bv,Ryu:2006ef}.
In \cite{Maldacena:2017axo}, a reduction of $S_R$ was interpreted as reflecting the second half of the Page curve for the original black hole \cite{Page:1993wv}, corresponding to the stage of the evaporation process beyond the Page time, which is indeed the appropriate stage for the black hole in our set-up.

The reduction in $S_R$ also reduces the traversability of the wormhole, as measured by $N$, since the Hayden-Preskill or traversable wormhole protocol relies crucially on a large amount of entanglement between the black hole and its early radiation \cite{Hayden:2007cs,Gao:2016bin,Maldacena:2017axo}.
This provides another perspective on the upper bound on $N$: if already $\mathcal{O}(1)$ bits destroy most of the entanglement, it is not possible to send any more information.
We will come back to this connection between the traversability of the wormhole and the entanglement entropy of one asymptotic boundary in the next section, when we consider sending multiple messages through the wormhole.

Let us finally comment on the increase of $R_2$.
As recently argued, this may be understood as a rejuvenation of the original black hole \cite{Almheiri:2019psf,Penington:2019npb}.
More precisely, assuming that the message has a much smaller entanglement entropy than the increase in black hole entropy, for example when the message is described by a pure state, this increase implies that the Page time for the new black hole of radius $R_2$ is (much) later than the original Page time \cite{Penington:2019npb}.
This rejuvenation may seem to be in tension with our observation above that $S_R$ is decreasing, which is suggestive of an old black hole instead.
However, as noted in \cite{Almheiri:2019psf}, the increased black hole entropy becomes available to the system only after a scrambling time.
Therefore, the increase in $R_2$ is compatible with the decrease of $R_3$ \textit{before} a scrambling time.
To compute the scrambling time and to see the subsequent increase in $S_R$ associated to the rejuvenation, as recently obtained in the context of JT gravity in \cite{Almheiri:2019psf}, one presumably has to go beyond our simple analysis and include quantum effects.
Nevertheless, we find it remarkable that our geometrical model is able to capture some of the features of black hole evaporation that were obtained in \cite{Maldacena:2017axo,Almheiri:2019psf,Penington:2019npb} using quantum field theoretic methods.

\subsection{NP interaction as model for black hole evaporation}\label{sec:bh-evap}

In the previous sections, we have discussed how the NP shockwave interaction results in a reduction of the entanglement entropy of the radiation subsystem and an increase in the black hole entropy.
We argued that this is consistent with the second half of the Page curve before a scrambling time.
In this section, we would like to refine our model in order to explore the usefulness of our set-up as a toy model for black hole evaporation.
In particular, we argue that the time-dependence of the traversibility of the wormhole, defined through $N$ in \eqref{eq:N-def}, in our setup is closely related to the time-dependence of the entanglement entropy of the radiation for an evaporating black hole.
The usefulness of this connection is that the definition of $N$ is much simpler than that of the latter, since it is computable purely from the backreacted geometry (in combination with the uncertainty principle).  
Furthermore, we will make use of the left-right symmetry of the extended black hole to reinterpret our set-up.
Indeed, we will now think about the right side as the actual black hole at the Page time, while the left side represents the early radiation.
In this case, the NP shockwave interaction provides a way of extracting energy from the old black hole and carrying it to the asymptotic boundary in Region I.
In particular, we propose to think of the passage of the positive shock(s) through the wormhole as late Hawking radiation of the black hole in Region I.
From this perspective, we again speculate on connections between observables computable from our geometry and the Page curve.

First of all, we note that if our set-up is to provide some model for black hole evaporation, it is very unnatural to allow the positive shockwave energy to saturate the bound \eqref{eq:bound-on-bits}.
Indeed, if the bound is saturated the NP interaction results in an instantaneous reduction of $R_3$.
To reflect the evaporation process more accurately, we impose that $S_R$, the entanglement entropy of the radiation subsystem, follows the Page curve. 
After the Page time, $S_R$ can be argued to follow the evolution of the Bekenstein-Hawking entropy of the remaining black hole \cite{Page:1993wv}.
This evolution can be determined as follows.
In three dimensions, the energy flux in thermal radiation is proportional to $2\pi RT^3$.
Using that for the BTZ black hole $T\sim R$ and $M\sim R^2$, one arrives at \cite{Horowitz:1993jc}:
\begin{equation}\label{eq:rate-evap-btz}
\frac{dM}{dt}=- cM^2 \qquad \Rightarrow \qquad M(t)=\frac{M_0}{M_0(t-t_0)+1}, 
\end{equation} 
with $M_0$ the initial mass and $t_0$ the starting time of the evaporation process, and we set the numerical constant $c=1$ which depends on details of the radiation.
In addition, note that $t$ represents the ordinary AdS-Schwarzschild time coordinate.
From the rate, it follows that BTZ black holes only evaporate asymptotically.

Given this rate, we propose to model the evaporation process as follows.
We will send multiple positive shockwaves from the left region at the integer times:
\begin{equation}
x_t\equiv a_t/|\alpha|=e^{t+t_0},\qquad t_0<0 \quad \& \quad t=1,\ldots, |t_0|\ .
\end{equation}
The $x_t$ are separated exponentially because they are measured in Kruskal time, whereas the rate is defined with respect to the AdS-Schwarzschild time $t$.
Note that the shocks are always sent in the traversable window, i.e. $0< a_t/|\alpha| \leq 1$.
Furthermore, we require that the shocks backreact on the geometry in such a way that after the shock, $R_3$ (proportional to $S_R$) changes according to the rate:
\begin{equation}\label{eq:r3t-rate}
R^{(t)}_3=\ell\sqrt{8G_NM_0\over M_0\log x_t+1}\ .
\end{equation}
This relation fixes the amount of energy $\beta_t$ of the shockwave, and therefore also the amount of information $N_t$ carried by the shockwave via the definition \eqref{eq:N-def} and the uncertainty relation \eqref{eq:uncertainty-relation}.

The explicit backreacted geometry for multiple shocks is constructed in Appendix \ref{sec:recursion}. It amounts to a set of recursion equations:

\medskip\noindent
(1) $R_2$-recursion:
\be\label{sec4:R2-recursion}
{R^{(i+1)}_2\left(R^{(i+1)}_2-R^{(i)}_2+(R^{(i+1)}_2+R^{(i)}_2)\beta_{i} A_{i+1}\right)\over R^{(i)}_2\left(R^{(i+1)}_2+R^{(i)}_2+(R^{(i+1)}_2-R^{(i)}_2)\beta_{i} A_{i+1}\right)}= A_{i+1}(\beta_{i}+\beta_{i+1})\ .
\ee
\noindent
(2) $R_3$-recursion:
\begin{align}\label{sec4:R3-recursion}
\hspace{-.4cm}(R^{(i+1)}_3)^2
&=\frac{(1-\beta_iA_{i+1})^2\left((R^{(i)}_2)^2-(R^{(i+1)}_2)^2-(R^{(i)}_3)^2\right)
+(1+\beta_iA_{i+1})^2\left({R^{(i+1)}_2R^{(i)}_3\over R^{(i)}_2}\right)^2}{4\beta_iA_{i+1}}.
\end{align}
\noindent
(3) $A$-recursion:
\be\label{sec4:A-recursion}
A_{i+1}=a_{i+1}-a_i+A_i{R^{(i-1)}_2\over R^{(i)}_2}\qquad\mbox{with}\qquad A_1=a_1,\quad R_2^{(0)}=R
\ee
where $R_2^{(i)}$ and $R_3^{(i)}$ are the horizon radii in the $i$-th subregion separated by the $i^{\rm th}$ and $(i+1)^{\rm th}$ P-shocks, which are sent at $V=a_i$ and $V=a_{i+1}$ respectively. 
$R_3^{(1)}$ is equal to $R_3$ in \eqref{BHmassredTraversable}.
$A_i$ is defined by $A_i\equiv V^{(i-1)}_2(a_{i})$ which is the location $a_i$ measured in the $(i-1)^{\rm th}$ $V$-coordinate in Region 2.

\begin{figure}
\centering
\begin{subfigure}{.5\textwidth}
  \centering
  \includegraphics[width=0.92\linewidth]{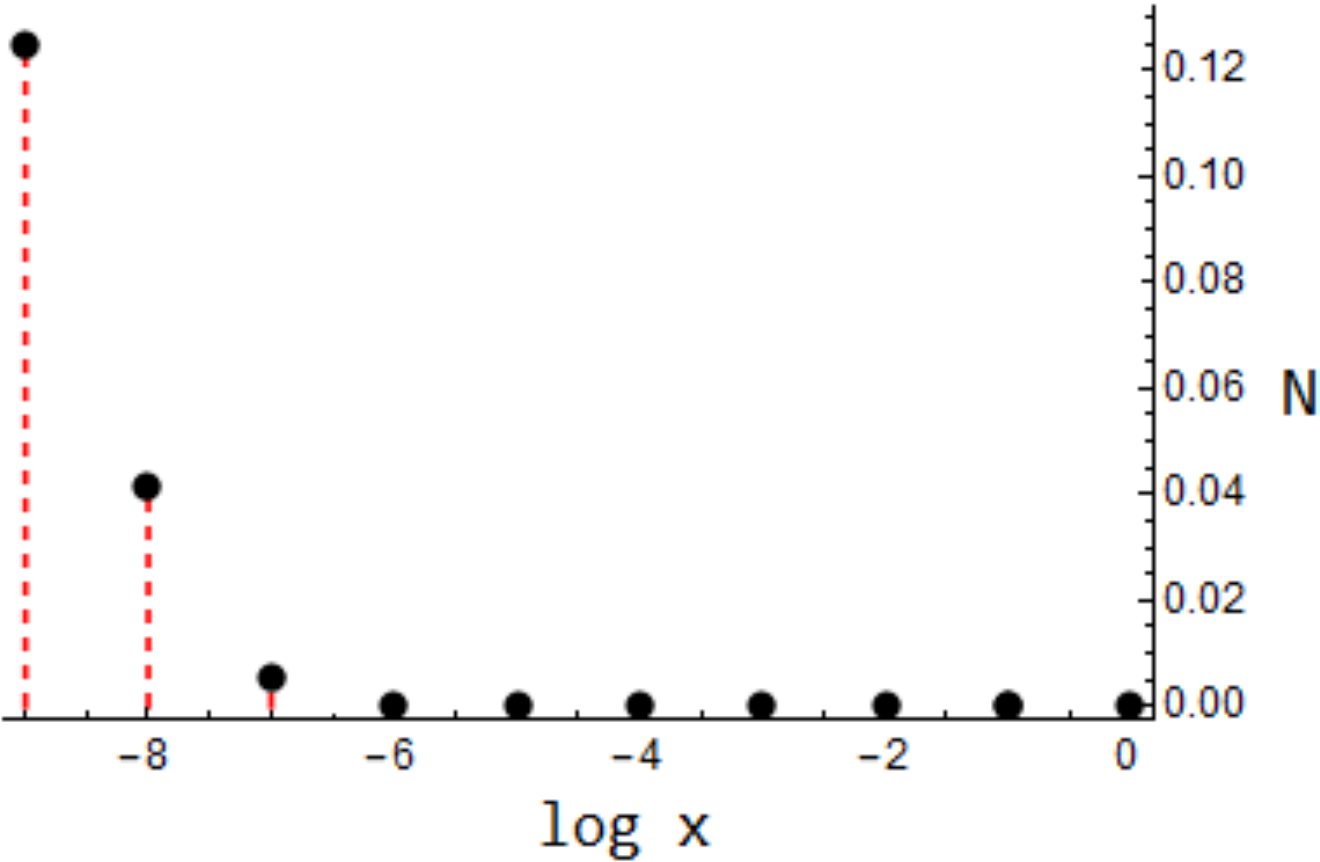}
  \caption{\textit{Plot of $N$ against $\log x$.}}
  \label{fig:N-x-sub}
\end{subfigure}%
\begin{subfigure}{.5\textwidth}
  \centering
  \includegraphics[width=1.0\linewidth]{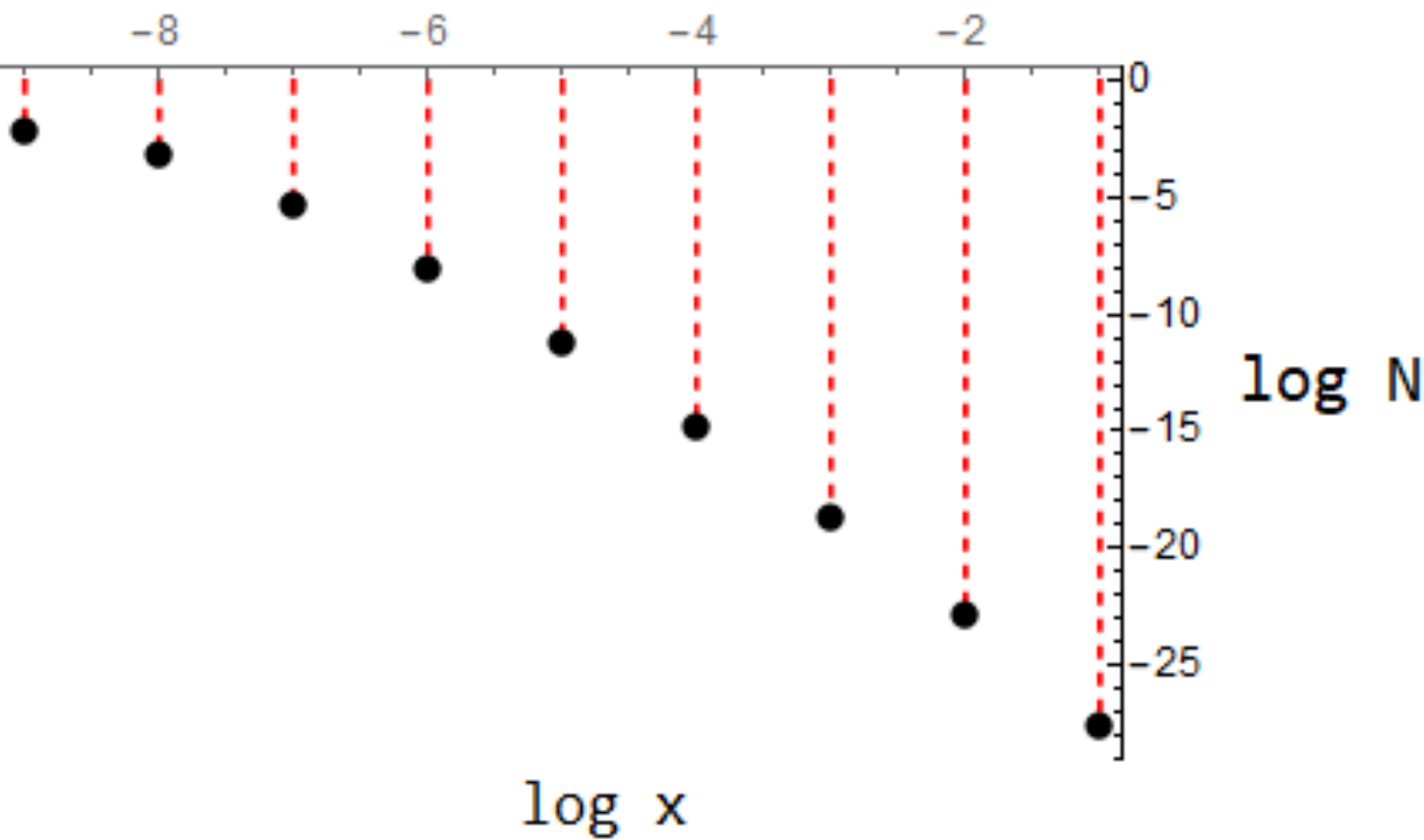}
  \caption{\textit{Plot of $\log N$ against $\log x$.}}
  \label{fig:log-N-x-sub}
\end{subfigure}
\caption{\textit{The plots show the behaviour of the amount of (qu)bits $N$ contained in the multiple shockwaves given the rate of evaporation \eqref{eq:rate-evap-btz}.}}
\label{fig:N-x}
\end{figure}

\begin{figure}
\centering
\begin{subfigure}{.5\textwidth}
  \centering
  \includegraphics[width=0.96\linewidth]{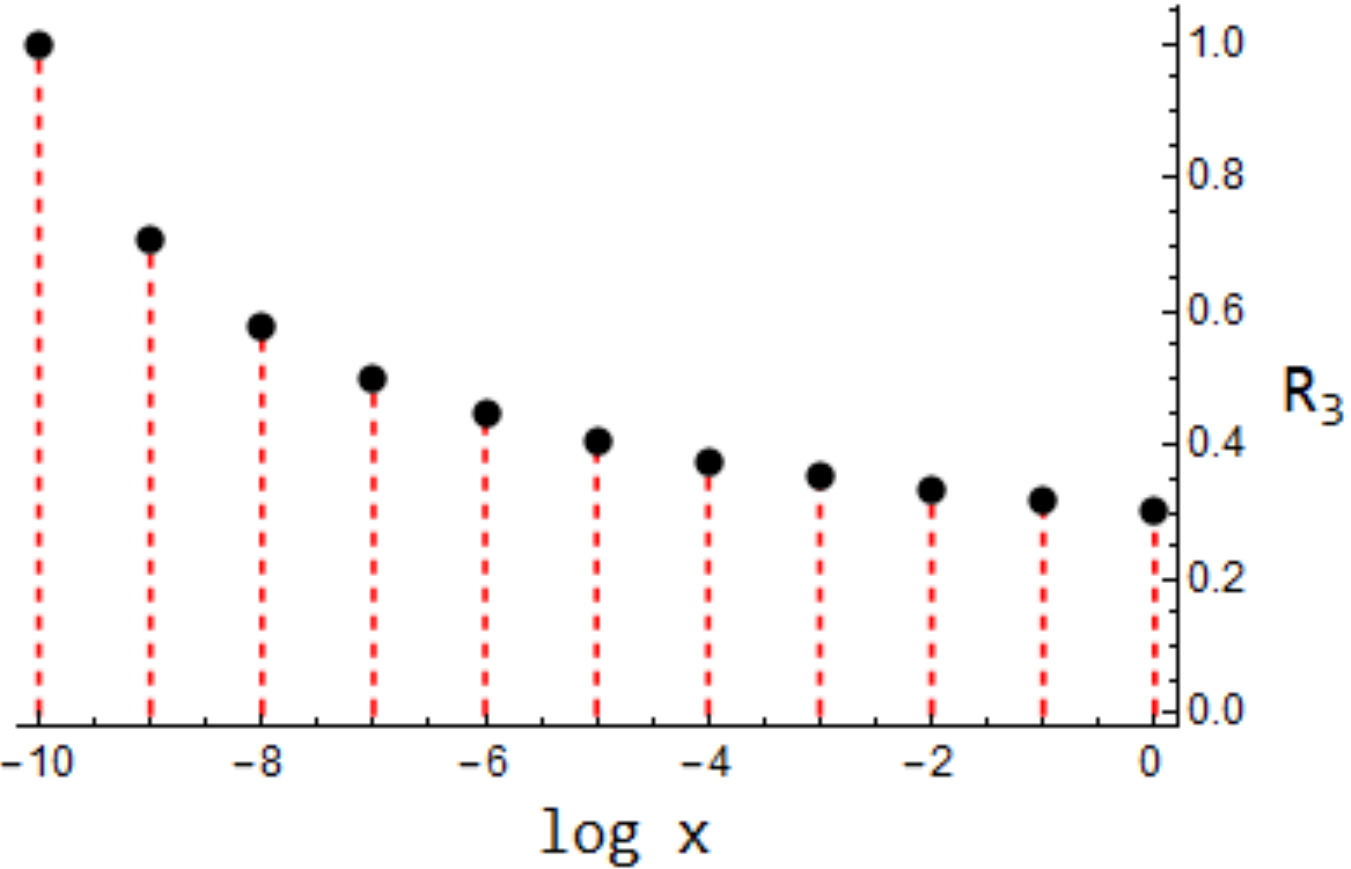}
  \caption{Plot of $R_3$ against $\log x$.}
  \label{fig:R3-x-sub}
\end{subfigure}%
\begin{subfigure}{.5\textwidth}
  \centering
  \includegraphics[width=0.96\linewidth]{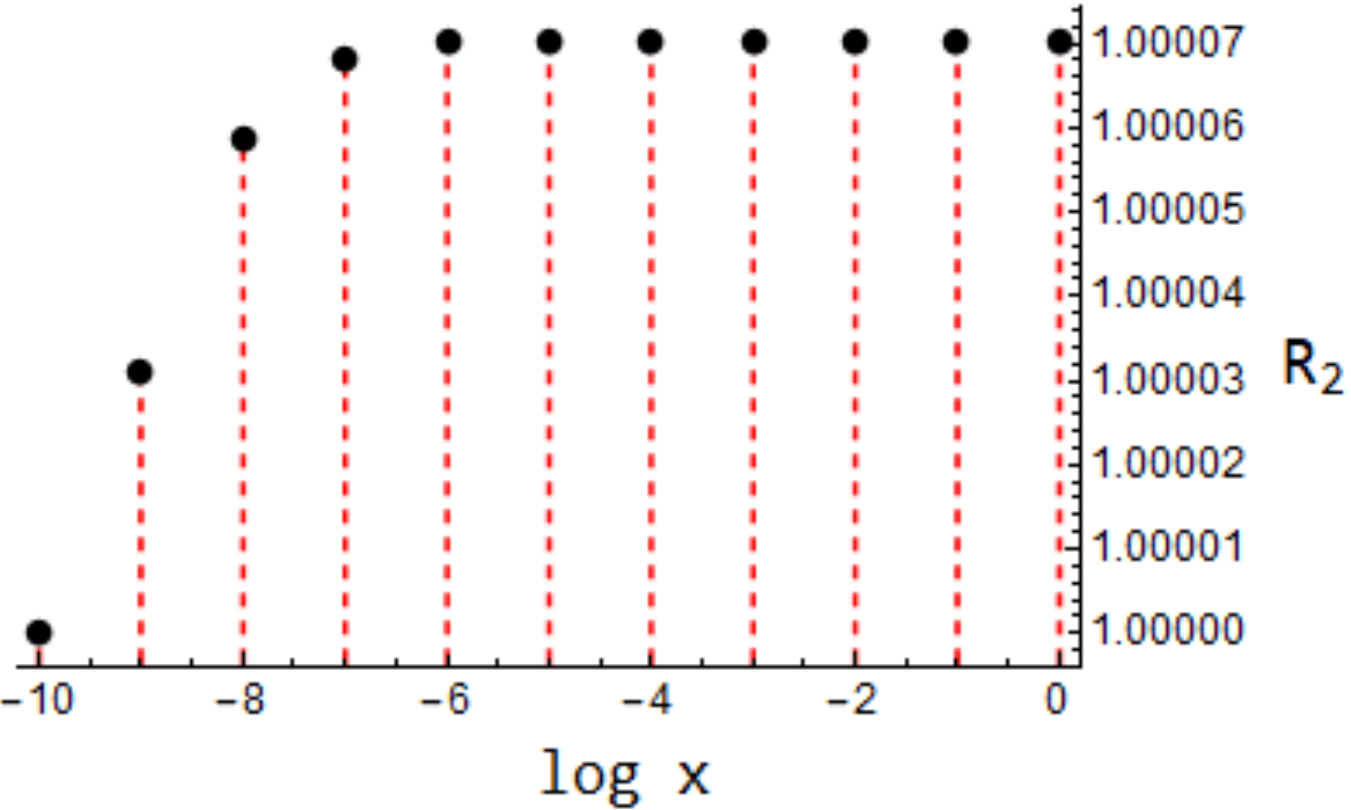}
  \caption{Plot of $R_2$ against $\log x$.}
  \label{fig:R2-x-sub}
\end{subfigure}
\caption{\textit{The plots show the behaviour for the left and right BTZ radii as a function of time. Note that $R_3$ decreases, by construction, whereas $R_2$ increases. Since the Bekenstein-Hawking entropy is $S=\pi R/(2G_N)$, these also represent the time-dependence of the black hole entropy in Region I and II. }}
\label{fig:R-x}
\end{figure}

We solved these equations numerically. 
Here, we will content ourselves by quoting  the results for the specific example of ten shockwaves with an initial time of $t_0=-10$.
We have plotted the values of the various $N_t$ in Figure \ref{fig:N-x} and the corresponding decrease in $R^{(t)}_3$ (radius/entropy in Region I) and increase $R^{(t)}_2$ (radius/entropy in Region II) in Figure \ref{fig:R-x}. 

Figure \ref{fig:N-x} shows essentially the same result as in the previous section: $N_t$ is monotonically decreasing.
A monotonic decrease of $N$ was to be expected from the discussion in Section \ref{sec:bound}, where we have seen that the uncertainty effect, however slightly, will dominate the boost effect at all times.
Since the physical rate of evaporation is largest at early times, the early shocks reduce $R_3$ most significantly, causing the steep decrease of $N$ in Figure \ref{fig:N-x-sub}.
That is, the early shocks reduce the traversability almost completely.

From the recursion relations \eqref{sec4:R2-recursion}, \eqref{sec4:R3-recursion} and \eqref{sec4:A-recursion}, one can check that the total amount of information in the multiple shocks is given by:
\begin{equation}
N_{\rm tot}C=\sum N_tC\approx 0.17
\end{equation}
Note that this is smaller than the upper bound on a single shock wave, which can be read from \eqref{eq:Nbound-numbers} to be $0.25$. 
This is of course as it should be.
In fact, we can say a little more: the linearity of shockwaves and the slow decrease of the upper bound on a single shockwave imply that if the black hole mass was fully reduced at early enough times in the multiple shock case, one would expect the energies of those shocks to sum up to $\lesssim 0.25$.\footnote{In the case of coincident shockwaves shot at $x=a=0$, the transferable information $N_{\rm tot}$ would saturate the inequality. This can be checked explicitly from the recursion relations \eqref{R2-recursion} and \eqref{R3-recursion}, which for coincident shocks at $x=0$ reduce for any of the shocks to the consistency condition of a single shockwave. In particular, this implies that the maximal amount of information sent through multiple coincident shocks is the same as the maximal information contained in a single shock wave at $x=0$, as expected from the linearity of superposition of shockwaves.}
However, only about $70\%$ of the black hole radius is reduced in the process at hand, which is consistent with the fact that $0.17$ is roughly $70\%$ of $0.25$.

We would now like to come back to the interpretation of our model as a toy model for black hole evaporation.
Figure \ref{fig:R-x} shows both the gradual reduction of $R_3$ and in the increase in $R_2$ in the multiple shock process.
In particular, the curve for $R_3$ now represents by construction the second half of the Page curve, as imposed by the evaporation rate \eqref{eq:rate-evap-btz}.
On the other hand, we see a slight increase of $R_2$ with subsequent shocks.
As argued at the end of Section \ref{sec:bound}, this increase leads to a rejuvenation of the original black hole after a scrambling time.
Therefore, after a scrambling time one expects that in a full quantum mechanical model $S_R$, i.e. $R_3$, should start to increase again.
This should subsequently improve the traversability of the wormhole in the sense that the amount of entanglement between the two sides is growing.
We interpret the fact that we only see a monotonic decrease of $N$, illustrated in Figure \ref{fig:N-x}, as being related to the fact that the rejuvenation of the black hole does not compete with the decrease in $S_R$.
Indeed, it can be seen from Figure \ref{fig:R-x} that the increase of $R_2$ is only a tiny fraction of the decrease in $R_3$.

On the other hand, we would expect that in a model where the initial evaporation is slow, such that the decrease of $R_3$ is limited at early times, the traversability could increase at some intermediate time.\footnote{Black holes that evaporate at a slower rate at early times exist; they are the negative specific heat black holes, such as small AdS black holes or asymptotically flat black holes, as opposed to the large BTZ black hole we consider here.}
Understanding traversability as entanglement between both sides, it appears that our model in this case could even reflect the rejuvenation of the black hole by the positive shockwaves.
We will further comment on this in Section \ref{sec:disc}.
  
\begin{figure}
\centering
  \includegraphics[width=0.6\linewidth]{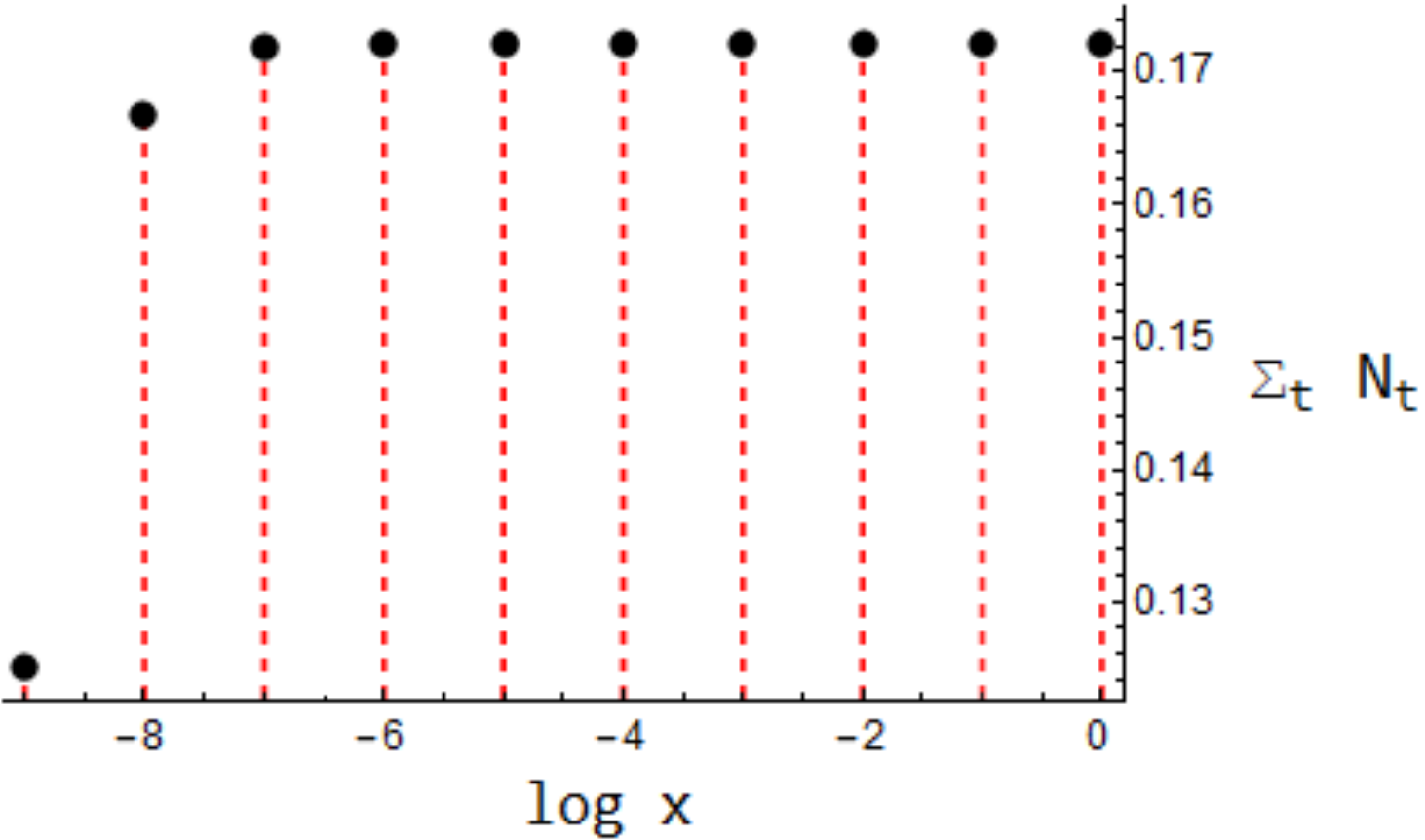}
\caption{\textit{The plot shows the plateauing behaviour of the cumulative information in the shocks. }}
\label{fig:cum-N-x}
\end{figure}

We conclude this section by reversing the roles of the early radiation and the old black hole in Figure \ref{fig:NP-shock-penrose}.
In this case, the black hole resides in Region I whereas the early radiation is represented by Region II.
In this interpretation, the black hole entropy $S=\pi R_3/(2G_N)$ is reduced by the shockwaves according to the physical rate.
Hence, our multiple shock geometry provides a concrete model for ``stimulated'' black hole evaporation, where the positive energy shockwaves play the role of late Hawking radiation.
An important feature of the traversibility $N$ is its drop as time progresses, as already commented on and illustrated in Figure \ref{fig:N-x}.
If we interpret this as the statement that most of the information leaks out of the black hole at early times after the Page time, we reproduce another qualitative aspect of the BTZ evaporation process.
Indeed, if one repeats the analysis of Page \cite{Page:2013dx} on the evolution of ``information'' in the radiation for the BTZ black hole, one finds that this information essentially plateaus briefly after the Page time.
This property is due to the positive specific heat of the BTZ under consideration, which implies that at late times there is hardly any Hawking radiation, i.e. information, coming out of the black hole.
Comparing this with the plateauing we find in the cumulative information contained in the shocks, 
\begin{equation}
N_i=\sum^{t=i}_{t=0} N_t,
\end{equation}
as illustrated in Figure \ref{fig:cum-N-x}, we indeed see a similar time-dependence.

Finally, we should admit that at the moment it is not clear to us how to think of the increase in $R_2$, which in this case is interpreted as an increase in the entanglement entropy of the radiation subsystem.

\subsection{Delay of the scrambling time}\label{sec:scrambling}

In our NP-shock model, the P-shock as a message carrier cruises through the wormhole from one side to the other without encountering any dramatic events. 
As suggested in \cite{Maldacena:2017axo}, the smooth ride of the message carrier may indicate that traversable wormholes are fast decoders. 
In other words, the time evolution in the context of a traversable wormhole is not a random unitary and the associated scrambling is weaker for appropriately timed messages. 


To understand this point better, we calculate the scrambling time of the NP-shock geometry in order to measure how fast or slow the traversable wormhole geometry scrambles the information. 
More precisely, we wish to see how traversability, i.e. the N-shock, affects the scrambling time of the BTZ black hole.
For this purpose, we follow the method developed in \cite{Shenker:2013pqa} using mutual information.
In this consideration, we view the P-shock as a consequence of a small perturbation with energy $E$ at the boundary of the BTZ black hole which was sent at a very early boundary time $t_w$. Due to a large boost, the energy measured by a local observer at the $t=0$ slice is highly blue-shifted to  $E_p= E\ell/R\exp(Rt_w/\ell^2)$ which creates a P-shock with $\beta=E/(4M_0)\exp(Rt_w/\ell^2)$ \cite{Shenker:2013pqa}.
Following \cite{Shenker:2013pqa}, we first compute the mutual information for the backreacted spacetime.
Mutual information is defined by:
\begin{equation}\label{eq:mutual-info}
I(A; B)=S_A+S_B-S_{A\cup B},
\end{equation}
where $A$ and $B$ are the spatial intervals in the left and right boundary CFTs, respectively, which define the entangling regions.
Moreover, $S_R$ denotes the entanglement entropy of the reduced density matrix associated to the boundary region $R$.
Holographically, via the Ryu-Takayanagi proposal \cite{Ryu:2006bv,Ryu:2006ef}, $S_A$ and $S_B$ are the lengths of geodesics connecting the two endpoints of the interval $A$ and $B$, respectively, whereas $S_{A\cup B}$ is the length of the minimum of two geodesics: (1) the disconnected one which is a simple sum of $S_A+S_B$ or (2) the connected one which is twice the geodesic length connecting the endpoints of the intervals $A$ and $B$.

\begin{figure}[t]
	\centering
	\includegraphics[ trim=3cm 14cm 7cm 3.5cm, width=0.5\textwidth]{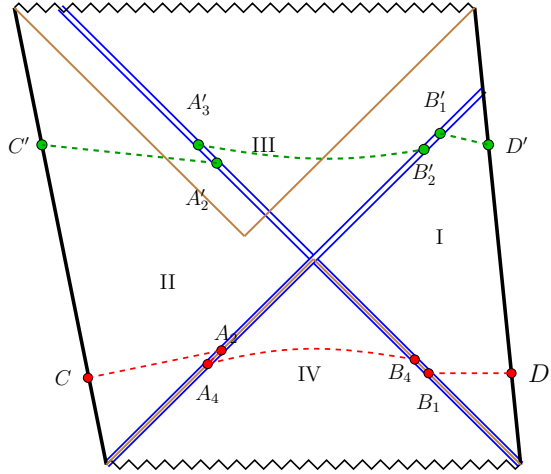}
	\caption{\small\it The Penrose diagram of BTZ black hole with the NP-shock wave in the case that both shocks  are shot along the horizon $UV=0$, i.e. $a=0$ in the geometry \eqref{eq:BTZpatchsolution}. The event horizons are drawn in brown and the shock waves in blue double lines. The points $C$ and $D$ are the boundary points with time lapse $t_L$ and $t_R$, respectively.  We want to find the shortest distance between $CD$ passing through $A$ and $B$ and crossing the horizons. The geodesics are shifted as they cross the shock waves. The subscripts denote the regions defined in \eqref{regions}. The geodesic $CABD$ passes through Region 4, while the geodesic $C'A'B'D'$ passes through Region 3. 
	}
	\label{fig:BTZpenrose2}
\end{figure}

In our patchwise geometry \eqref{eq:BTZpatchsolution}, there are two possible geodesics connecting the two boundary points in Figure \ref{fig:BTZpenrose2}. One possibility is the geodesic $CABD$ which passes through Region 4. We expect this to be the relevant geodesic if the P-shock is shot at sufficiently early times. 
Since the metric is locally BTZ in each patch, the geodesics can be easily calculated by using embedding coordinates:
\begin{equation}
\cosh \frac{d}{\ell} = T_1 T_1'+T_2T_2' -X_1 X_1'-X_2 X_2',
\end{equation}
where the relation of the embedding coordinates $(T_1, T_2, X_1, X_2)$ to the Kruskal and Schwarzschild coordinates are given in \eqref{eq:coordinates} in Appendix \ref{sec:BTZmetricscoordinate}.
Note that our $(U, V)$ coordinates are interchanged relative to those in \cite{Shenker:2013pqa}, and the Penrose diagram in Figure \ref{fig:BTZpenrose2} should also be understood in this convention. 

For simplicity and to highlight only the essential point, we consider the geodesics on the background \eqref{eq:BTZpatchsolution} in the $a=0$ limit.
In this case, the coordinates in Region 2 become $(U_2,V_2)=(U+\beta, V)$. The horizon radius $R_2=R$, as can be seen from \eqref{eq:regionIImassbyshock}. 
Let us denote the geodesic length in each patch by $d_i$. We first consider the geodesics passing through Region 4 which intersect with the horizons at $A(U,0)$ and $B(0,V)$ when expressed in the Region 4 coordinates. In terms of the coordinates in Region 2, these points are expressed by $A(U+\beta,0)$ and $B(0,V)$. Similarly, the point $B$ is expressed by $B(0,V+\alpha)$ in Region 1. Then, the geodesic passing through $A(U,0)$ and $B(0,V)$ has length 
\begin{eqnarray}
\cosh \frac{d_1}{\ell} &=& \frac{r}{R} +\frac{V+\alpha}{R}\sqrt{r^2-R^2} e^{\frac{R}{\ell^2}t_R}
\\ 
\cosh \frac{d_2}{\ell} &=& \frac{r}{R} + \frac{U+\beta}{R}\sqrt{r^2-R^2} e^{-\frac{R}{\ell^2}t_L}   \\
\cosh \frac{d_4}{\ell} &=& 2UV+1
\end{eqnarray}
in each patch, respectively. By solving the saddle point equations for $U$ and $V$, we find that the geodesic length at large $r$ is 
\begin{equation}\label{eq:lengthregionIV}
\frac{d}{\ell} \sim 2 \ln \frac{r}{R} +2\ln \left[ 2\cosh \left(\frac{R}{2\ell^2}(t_R-t_L)\right) +\alpha e^{\frac{R}{2\ell^2}(t_L+t_R)} +\beta e^{-\frac{R}{2\ell^2}(t_L+t_R)}  +\alpha \beta e^{\frac{R}{2\ell^2}(t_R-t_L)}\right]\ .
\end{equation}
The first term is the standard UV divergent piece near the boundaries. For the finite part, we can recover the single shock wave result in \cite{Shenker:2013pqa} by setting one of the shocks to zero.
In the equal time slice at $t_L=t_R=0$, this result simplifies to 
\begin{equation}\label{eq:lengthregionIVequalt=0}
\frac{d}{\ell} \sim 2 \ln \frac{r}{R} +2\ln \big( 2 +\alpha  +\beta   +\alpha \beta\big)\ .
\end{equation}

The scrambling time is determined by the zero of the mutual information $I(A; B)$ (see \eqref{eq:mutual-info}) by setting $\beta \sim E/(4M_0)e^{Rt_w/\ell^2}$, that is, the time $t_w=t_*$ at which the disconnected Ryu-Takayanagi surface \cite{Ryu:2006bv,Ryu:2006ef} starts dominating over the connected one.
One can now see that the effect of the N-shock is to rescale $\beta$ in \eqref{eq:lengthregionIVequalt=0} by a factor of $1-|\alpha|$. Using the result of the mutual information in \cite{Shenker:2013pqa}, the scrambling time of our traversable wormhole can be determined: 
\begin{equation}\label{eq:delayscrambletime}
t_* = \frac{\beta_{H}}{2\pi} \ln \frac{S}{1-|\alpha|},
\end{equation}
where $\beta_H$ and $S$ are the inverse Hawking temperature and the Bekenstein-Hawking entropy of the initial BTZ black hole, respectively. Therefore, we have demonstrated that the scrambling time is delayed by the presence of the negative shock wave. 
As it takes longer to scramble or ramdomize information in traversable wormholes compared to the non-traversable ones, it becomes easier to decode the information.
Put differently, the information can, in principle, be delivered before it gets completely scrambled. This is consistent with the fact that the message can be sent through the traversable wormhole without experiencing much disturbance.

We can also perform similar calculations in Region 3. The geodesics in Region 3 pass through $C'A'B'D'$. It is more complicated because even in the $a=0$ limit, the $(U_3, V_3)$ coordinates are not given by simple expressions:
\begin{equation}
\quad (U_3, V_3)=\left({R_3-R+(R_3+R)\alpha U\over c\left(R_3+R+(R_3-R)\alpha U\right)}, {R_3-R+(R_3+R)\beta V\over \gamma\left(R_3+R+(R_3-R)\beta V\right)}\right)
\end{equation}
where $c= V_3(\alpha)$ and $\gamma=U_3(\beta)$ and $c\gamma=(R_3-R)/(R_3+R)$. 
The geodesic length in each patch is found to be:
\begin{eqnarray} \nonumber
y_1 &=& \frac{r}{R} \frac{1- \alpha U}{1+\alpha U} -\frac{U}{1+\alpha U} \frac{\sqrt{r^2-R^2}}{R} \exp\left(-\frac{Rt_R}{\ell^2}\right) +\frac{\alpha}{1+\alpha U} \frac{\sqrt{r^2-R^2}}{R} \exp\left(\frac{Rt_R}{\ell^2}\right ) \\ \nonumber
y_2 &=& \frac{r}{R} \frac{1-\beta V}{1+\beta V} -\frac{V}{1+\beta V} \frac{\sqrt{r^2-R^2}}{R} \exp\left(\frac{Rt_L}{\ell^2}\right) +\frac{\beta}{1+\beta V} \frac{\sqrt{r^2-R^2}}{R} \exp\left(-\frac{Rt_L}{\ell^2}\right)   \\
y_3 &=& \frac{1+\alpha U+\beta V+2UV +\alpha\beta UV}{(1+\alpha U)(1+\beta V)}
\end{eqnarray}
where $y_i=\cosh(d_i/\ell)$. Solving the saddle point equations for $U$ and $V$, we find that the geodesic length at large $r$ is  
\begin{eqnarray}\nonumber
\frac{d}{\ell} &\sim& 2 \ln \frac{r}{R} +2\ln \left[ 2\cosh \left(\frac{R}{2\ell^2}(t_R-t_L)\right) +\alpha e^{\frac{R}{2\ell^2}(t_L+t_R)} +\beta e^{-\frac{R}{2\ell^2}(t_L+t_R)}  +\alpha \beta e^{\frac{R}{2\ell^2}(t_R-t_L)}\right] \\ \label{eq:lengthregionIII}
&&+  2\ln\Big[ \alpha \beta +\alpha e^{\frac{R}{\ell^2}t_L} +\beta e^{-\frac{R}{\ell^2}t_R} +e^{\frac{R}{\ell^2}(t_L -t_R)}\Big]\ .
\end{eqnarray}
However, this is always greater than \eqref{eq:lengthregionIV} passing through Region 4. Thus the shortest distance is given by \eqref{eq:lengthregionIV}.


\section{Conclusion}\label{sec:disc}

We studied traversable BTZ wormholes in the shockwave limit as a realization of the Hayden-Preskill protocol of information recovery \cite{Hayden:2007cs}, building on the previous works  \cite{Gao:2016bin} and  \cite{Maldacena:2017axo}. The BTZ black holes can be rendered traversable by a negative energy shockwave. Following Gao, Jafferis and Wall \cite{Gao:2016bin}, we showed that the negative shock is dual to the infinite boost limit of a specific double trace deformation which couples the left and right CFTs. 

In order to study the mechanism of information transfer through the wormhole in-depth, we sent in a message via the traversable wormhole, treating the backreaction of the message as a positive energy shockwave. The corresponding spacetime is that of colliding spherical shells in the BTZ black hole, which we explicitly constructed using the patchwise construction developed in \cite{Dray:1984ha,Dray:1985yt}. 
The backreaction of the message results in a reduction of the traversability of the wormhole, and we found that the upper bound on the amount of (qu)bits that can be sent in a message is only of order $\mathcal{O}(1)$. Even though this is disappointing, we argued that it is consistent with the late-time results of \cite{Maldacena:2017axo}.
In fact, our geometrical analysis provides a particularly simple perspective on this result, and explicitly shows how gravitational backreaction reduces the traversibility of the wormhole.

We also observed that the NP shock geometry results in an increase of the black hole radius, $R_2$, whereas it decreases $R_3$, which we interpreted as (proportional to) the entanglement entropy of the Hawking radiation.
Interpreting the empty extended BTZ geometry as the geometrization of an evaporating BTZ black hole at the Page time, we argued that these features reflect the stage of black hole evaporation as described by the second half of the Page curve.
This observation motivated us to construct a multiple shockwave geometry and to study it as a toy model for black hole evaporation. 
We estimated the amount of information in each shock, which allowed us to relate the traversibility of the wormhole with the evaporation process of the black hole.
In particular, we found that traversibility is reduced as the evaporation process continues beyond the Page time, as expected from the reduction in entanglement between the black hole and its Hawking radiation.
We related these features to some of the results derived using quantum field theoretic methods in \cite{Maldacena:2017axo,Almheiri:2019psf,Penington:2019npb}.
 
Finally, we examined the idea that the traversable wormhole prevents the infalling message from being fully randomized and provided evidence for this by showing that the scrambling time for the BTZ black hole is delayed by the presence of the negative shockwave.	
	
In our study of the CFT nonlocal coupling that creates a negative shockwave, the conformal dimensions of the left and right operators needed to be $\Delta=1/2$, which coincides with a special value noticed in \cite{Gao:2016bin} for the coupling applied during a finite time interval.
In particular, \cite{Gao:2016bin} noted that for a coupling with $\Delta > 1/2$ the stress-tensor is singular at the ends of the time interval but integrable in the sense that the averaged null energy remains finite. In fact, we found a similar singular behavior of the stress-tensor in the $\Delta > 1/2$ case. 
This is a true singularity in our case since our coupling is turned on only at an instant of time, similar to \cite{Maldacena:2017axo}, corresponding to the limit of a collapsed time interval. This limit singled out the $\Delta=1/2$ operators as the only ones that will lead to a non-singular coupling. The infinite boost limit, instead, required us to include a factor of $U_0$ in the coupling to keep it finite.
However, we would like to comment on the fine print of the $\Delta > 1/2$ case. A more rigorous treatment of \eqref{eq:vv-derivs-two-point-function-3} revealed that the singularity for the operators of $\Delta = 1/2+n$ with $n\in \mathbb{Z}_+$ turns out to be the $n$-th derivative of the $\delta$-function, $\delta^{(n)}(V)$. It would be very interesting to understand the type of geometry which generates this stress-tensor, and whether it can be related to a shockwave geometry.

We have focused our attention on the BTZ wormholes. However, it appears that much of our findings carry over to more general cases, not only to lower and higher dimensions, but also to asymptotically flat and de Sitter wormholes without putative field theory duals. In fact, negative energy shockwaves can be created by   the infinite boost limit of nonlocal {\it bulk} couplings,\footnote{A microscopic motivation for such non-local couplings is less clear in non-AdS spacetimes.} for example, in flat spacetime, in much the same way as in the AdS case. The spacetimes of colliding spherical shells can be similarly constructed for other neutral non-rotating black holes. The formula \eqref{BHmassredTraversable} for the black hole mass reduction, in particular, turns out to be of a rather universal form regardless of dimensions and asymptotics. Thus the mechanism of information transfer we spelled out in this work applies to a much wider class of traversable wormholes. This may not be surprising because the NP interaction, mediating the information transfer, is intrinsically (patchwise) local whereas the dimensionality and cosmological constant of the spacetime only play a secondary role. We will report more on the universality of this mechanism in the near future \cite{hirano2019}. 
	
Cases of particular interest concern black holes with a negative specific heat, such as asymptotically flat or small AdS black holes.
We expect that if we are able to repeat the analysis of Section \ref{sec:bh-evap} for these black holes, we might see more clearly how these black holes rejuvenate due to the positive energy shocks.
In particular, we expect the traversability of the associated wormholes to increase at early enough times, as opposed to the monotonic decrease observed for the large BTZ black hole in Figure \ref{fig:N-x}. 
Relating traversability to the entanglement between the black hole and its radiation, we arrive at an increase of the entanglement entropy. 
This is indicative of a rejuvenation of the black hole, corresponding to the first part of the Page curve (see Figure \ref{fig:PageCurves-intro} for examples of Page curves).

The traversable wormhole protocol for information recovery apparently hinges on the thermofield double (TFD) state. It is an important question whether or not this protocol could still work without it. One way to address this question may be to consider the NP-shock model in Vaidya-AdS spacetime. In this case, suppose Alice loses her diary at some time on the boundary before the collapsing shell starts off. Then, if the collapsing shell is sent into the bulk at a time not much later than at which the diary was lost, the diary will be trapped in the newly formed black hole.
However, Bob could send a negative energy shockwave into the bulk, perhaps by introducing a non-local coupling in the spirit of \cite{deBoer:2019kyr}, such that the diary could be retrieved from the black hole before it hits the singularity.
The mechanism underlying this retrieval is of course again the traversable wormhole protocol.	
	
The simplicity of our model makes the workings of the traversable wormhole protocol explicit and transparent and allows us to carry out in-depth studies. However, as we have seen in Sections \ref{sec:bound} and \ref{sec:bh-evap}, one drawback is that it is a rather ineffective example of the information recovery protocol: we cannot send more than a few (qu)bits. This is due to the fact that the message and coupling have a large relative boost and the consequent backreaction of the message spoils the traversability. However, there is an obvious way to improve the efficacy of information transfer. As remarked in Section \ref{sec:bound}, we can send a negative shockwave away from the (initial) black hole horizon, i.e. at $U\ne 0$ in our convention. In this case, the relative boost between the positive and negative shockwaves becomes smaller. Thus we expect that there will be a significant improvement in the capacity of information transfer, even though the positive shockwave still backreacts to weaken the traversability. Also, we should emphasize our analysis does not rule out other mechanisms which allow information transfer. As recently discussed in \cite{Bao:2019rjy}, the shock wave model is analogous to classical motion of a particle in the bulk geometry. The capability of information transfer could be improved if one considers a multiple wormhole geometry, which would be controlled by a quantum particle motion instead. It would be interesting to have a more detailed bulk description of the proposed model.

Finally, the black hole evaporation discussed in this paper is classically stimulated as opposed to quantum spontaneous. However, one can imagine that, if the pair creation picture of Hawking radiation is assumed, the negative energy particles behave as baby N-shocks and the positive ones as baby P-shocks. Then the microscopic version of the NP interaction may occur: a cloud of many virtual NP pairs is formed in the vicinity of the horizon and some of P-shocks inside are crossed by infalling N-shocks created on the outside.
Those P-shocks that are hit by N-shocks can then escape from the black hole as Hawking radiation.  
This picture is reminiscent of an idea suggested by Page a long time ago \cite{Page:1993up}.

\section*{Acknowledgements}

This work was supported in part by the National Research Foundation of South Africa and DST-NRF Centre of Excellence in Mathematical and Statistical Sciences (CoE-MaSS). YL thanks the support from the Department of Science and Technology and the National Research Foundation's South African Research Chairs Initiative for post-doctoral support.
SvL gratefully acknowledges support from the Simons Foundation Grant Award ID:509116.
Opinions expressed and conclusions arrived at are those of the authors and are not necessarily to be attributed to the NRF or the CoE-MaSS.

\appendix
\section{Coordinate systems for BTZ black holes} \label{sec:BTZmetricscoordinate}

The (neutral) BTZ black holes are locally AdS$_3$ \cite{Banados:1992wn}. The latter is a hyperboloid defined by
\be
T_1^2+T_2^2-X_1^2-X_2^2=\ell^2,
\ee
where $\ell$ is the radius of the AdS$_3$ space. In this paper we work in the two coordinate systems, the Kruskal coordinates $(U, V, \phi)$ and the Schwarzschild coordinates $(\tau, r, \phi)$ defined and related by
\begin{align}\label{eq:coordinates}
\begin{split}
T_1 &= \frac{V+U}{1+UV} = \frac{1}{R} \sqrt{r^2-R^2} \sinh \frac{R \tau}{\ell^2}\\ 
T_2 &= \frac{1-UV}{1+UV} \cosh \frac{R\phi}{\ell} = \frac{r}{R} \cosh \frac{R\phi}{\ell} \\ 
X_1 &= \frac{U-V}{1+UV} =\frac{1}{R} \sqrt{r^2-R^2} \cosh \frac{R \tau}{\ell^2} \\
X_2 &= \frac{1-UV}{1+UV} \sinh \frac{R\phi}{\ell} =\frac{r}{R} \sinh \frac{R\phi}{\ell}.
\end{split}
\end{align}
Note that our $X_1$ is minus the oft-used convention. 
In Schwarzschild coordinates, the BTZ black hole has an event horizon at $r=R$ 
\begin{align}
ds^2=-{r^2-R^2\over \ell^2}d\tau^2+{\ell^2\over r^2-R^2}dr^2+r^2d\phi^2\ ,
\end{align}
where $\phi\sim \phi+2\pi$ and $\ell$ is the AdS radius.
This is a chart that covers the outside of the horizon and corresponds to an observer at an asymptotic boundary. 
To study the global structure of BTZ black holes, it is more useful to use the Kruskal coordinates in terms of which the metric takes the form 
\be
ds^2={-4\ell^2dUdV+R^2(1-UV)^2d\phi^2\over (1+UV)^2}\ .
\ee
At the right boundary $r\to\infty$, the two coordinate systems are related by
\be
{2U\over 1-UV}=e^{R\tau/\ell^2}\ ,\qquad {2V\over 1-UV}=-e^{-{R\tau/\ell^2}}\ .
\ee
which imply
\be\label{eq:UVtauboundary}
UV=-1\ ,\qquad\quad U=e^{R\tau/\ell^2}\ ,\qquad\quad V=-e^{-{R\tau/\ell^2}}\ .
\ee
The relation at the left boundary can be found by shifting the Schwarzschild time $\tau$ to $\tau +i\beta/2$ with $\beta=2\pi\ell^2/R$.

For use of Section \ref{sec:Wightman}, we show the identity \eqref{eq:identityBTZ}:
\begin{eqnarray} \nonumber
-{(r^2-R^2)^{1\over 2}\over R}\cosh\left(\frac{R}{\ell^2}(\tau-\tau_0)\right) &=& {(r^2-R^2)^{1\over 2}\over R}\left[ \sinh \frac{R\tau}{\ell^2} \sinh \frac{R \tau_0}{\ell^2} - \cosh \frac{R\tau}{\ell^2} \cosh \frac{R \tau_0}{\ell^2}\right] \\ \nonumber
&=&  {2(UV_0+VU_0)\over(1+UV)(1-U_0V_0)}\ .
\end{eqnarray}

\section{Recursive consistency conditions for multiple shocks}\label{sec:recursion}

In Section \ref{sec:patchmetric} we have constructed the spacetime of two colliding spherical shells in the BTZ black hole. In this appendix we provide some details of the construction of colliding multiple shock geometry which was studied in Section \ref{sec:bh-evap} in the context of black hole evaporation.
To be more precise, we consider an N-shock at $U=0$ as in the two-shock case, but we send an arbitrary number of multiple P-shocks of strengths $\beta_1, \beta_2, \beta_3, \cdots$ one after another at $V=a_1<a_2<a_3<\cdots$. 
This is the spacetime in which one negative shockwave is colliding with multiple positive shockwaves. 

Using the patchwise construction of Dray and 't Hooft \cite{Dray:1985yt, Shenker:2013pqa, Shenker:2013yza}, we obtain the following consistency conditions:

\noindent
{\bf (1) $A$-recursion:}
\be\label{A-recursion}
A_{i+1}=a_{i+1}-a_i+A_i{R^{(i-1)}_2\over R^{(i)}_2}\qquad\mbox{with}\qquad A_1=a_1,\quad R_2^{(0)}=R\ ,
\ee

\medskip\noindent
{\bf (2) $R_2$-recursion:}
\be\label{R2-recursion}
{R^{(i+1)}_2\left(R^{(i+1)}_2-R^{(i)}_2+(R^{(i+1)}_2+R^{(i)}_2)\beta_{i} A_{i+1}\right)\over R^{(i)}_2\left(R^{(i+1)}_2+R^{(i)}_2+(R^{(i+1)}_2-R^{(i)}_2)\beta_{i} A_{i+1}\right)}= A_{i+1}(\beta_{i}+\beta_{i+1})\ ,
\ee

\noindent
{\bf (3) $R_3$-recursion:}
\begin{align}\label{R3-recursion}
\hspace{-0cm}(R^{(i+1)}_3)^2
&={(1-\beta_iA_{i+1})^2\over 4\beta_iA_{i+1}}\left((R^{(i)}_2)^2-(R^{(i+1)}_2)^2-(R^{(i)}_3)^2\right)
+{(1+\beta_iA_{i+1})^2\over 4\beta_iA_{i+1}}\left({R^{(i+1)}_2R^{(i)}_3\over R^{(i)}_2}\right)^2\ ,
\end{align}
where $R_2^{(i)}$ and $R_3^{(i)}$ are the horizon radii in the $i$-th subregions separated by the $i$-th P-shock at $V=a_i$ within Regions 2 and 3, respectively. 
$R_3^{(1)}$ is equal to $R_3$ in \eqref{BHmassredTraversable}.
$A_i$ is defined by $A_i\equiv V^{(i-1)}_2(a_{i})$ where $V_2^{(0)}=V$ and $V_2^{(1)}=V_2=V-V_0$ in the two-shock case. Thus, for example, $A_2=a_2-a_1+a_1R/R_2=a_2-a_1+V_2(a_1)=V^{(1)}_2(a_2)$.

In section  \ref{sec:bh-evap} we have recursively found $\beta_{i}$ with the input $R_3^{(k)}=f_k R$ ($k\le i$) where $f_k$ defines the fraction in reference to the initial black hole radius $R$. For the process where the black hole gradually evaporates, one must require that $1\ge f_1>f_2>f_3>\cdots\ge 0$. The choice of the fraction $f_i$ is determined by the evaporation rate of our interest. Then imposing the uncertainty relation yields an upper bound on the number of (qu)bits $N_i$ in the $i$-th P-shock.

We are now going to derive the consistency conditions. It is straightforward but tedious. Since the consistency of the $(i+1)$-th P-shock only depends on its relation to the $i$-th data, we can restrict ourselves to the case of the second P-shock of strength $\beta'$ at $V=a'$ added to the first P-shock of strength $\beta$ at $V=a<a'$. Then the pattern repeats itself to be generalized to the generic $i$-th recursion.

The second P-shock creates the new subregions within Regions 2 and 3: In Region 2, one is a strip $\Delta 2$ $(U\le 0, a\le V\le a')$ and the rest of Region 2 is denoted by $2'$ $(U\le 0,  V\ge a')$.
Similarly, in Region 3, one is a strip $\Delta 3$ $(U\ge 0, a\le V\le a')$ and the rest of Region 3 is denoted by $3'$ $(U\ge 0, V\ge a')$. 

We need to patch (1) $\Delta 2$ and $2'$, (2) $\Delta 3$ and $3'$, and (3) $2'$ and $3'$. Since subregions $2'$ and $3'$ are determined by the first two patchings, the last one gives us the consistency conditions which we are after.

\paragraph{(1) Continuity between $\Delta 2$ \& $2'$:}

In much the same way as that in Section \ref{sec:patchmetric}, the first condition in the continuity along $V=a'$ reads
\begin{align}
U_{2'}(U_2)=&{R'_2-R_2+(R'_2+R_2)V_2(a')U_2\over V_{2'}(V_2(a'))\left(R'_2+R_2+(R'_2-R_2)V_2(a')U_2\right)}\ .
\end{align}
where $U_{2'}$ is as a function of $U_2$ and $V_{2'}$ as a function of $V_2$ which itself is a function of $V$.
Since the shift of the horizon at $U=0$ in Region $2'$ must be given by the {\it total} strength of the two P-shocks, we idenfity
\begin{align}
U_{2'}(U_2(0))=U_2(\beta)={R'_2-R_2+(R'_2+R_2)V_2(a')\beta\over V_{2'}(V_2(a'))\left(R'_2+R_2+(R'_2-R_2)V_2(a')\beta\right)}
= \beta+\beta'\ .
\end{align}
Combining this with the second condition yields
\be
V'_{2'}(V_2(a'))V_2(a')R_2=V_{2'}(V_2(a'))R'_2
\ee 
where $V'_{2'}=\partial_{V_2}V_{2'}$. We can consistently choose $V_{2'}(V_2)=V_2+V_{2'}(V_2(a'))-V_2(a')$. Then we have
\be\label{Arelation}
V_2(a')R_2=V_{2'}(V_2(a'))R'_2\ .
\ee
These conditions then require that
\be
{R_2'\left(R_2'-R_2+(R_2'+R_2)A\beta\right)\over R_2\left(R_2'+R_2+(R'_2-R_2)A\beta\right)}= A(\beta+\beta')\ .\label{R2prime}
\ee
where $A\equiv V_2(a')=a'-a+V_2(a)$.
This can be solved to 
\be
{R_2'\over R_2}=
\frac{1+A \beta' +A^2 \beta(\beta+\beta') +\sqrt{(1+A\beta'+A^2\beta  (\beta + \beta'))^2+4 A\left(1-A^2 \beta ^2\right) 
		(\beta + \beta')}}{2(1+ A\beta)}
\ee
which reduces to $R'_2=R_2$ when $\beta'=0$ as it should.

\paragraph{(2) Continuity between $\Delta 3$ \& $3'$:}

Similarly, the first condition of the continuity along $V=a'$ reads
\be\label{2PIII'}
U_{3'}(U_3)={R_3'-R_3+(R_3+R'_3)V_3(a')U_3 \over c'\left(R'_3+R_3+(R'_3-R_3)V_3(a') U_3\right)}\ ,
\ee
where we introduced $V_{3'}(V_3(a'))=c'$. $U_{3'}$ is a function of $U_3$ and $V_3$ here is regarded a function of $V$.
Combining this with the second condition yields
\be\label{eq:v3prime}
V_{3'}'(V_3(a'))V_3(a') R_3 = c' R'_3\ ,
\ee
where $V'_{3'}=\partial_{V_3}V_{3'}$.

\paragraph{(3) Continuity between $2'$ \& $3'$:}

The first condition of the $2'$-to-$3'$ continuity and the combined condition are given by
\begin{eqnarray}
&&V_{3'}(V_{2'})={R_3'-R_2'+(R'_3+R'_2)U_{2'}(\beta)V_{2'}\over \gamma'\left(R'_3+R_2'+(R'_3-R'_2)U_{2'}(\beta) V_{2'}\right)}\ ,\label{2PII'III'} \\
&& U_{3'}'(U_{2'}(\beta))U_{2'}(\beta)R'_2=\gamma' R'_3 \label{eq:u3prime}
\end{eqnarray}
where $U_{3'}'=\partial_{U_{2'}}U_{3'}$ and 
we introduced $\gamma'=U_{3'}(U_{2'}(\beta))$ and we used $U_2(0)=\beta$. $V_{3'}$ is regarded a function of $V_{2'}$ and $U_{3'}$ as a function of $U_{2'}$.

Now since the LHS of \eqref{2PIII'} and \eqref{2PII'III'} evaluated at $U=0$ and $V=a'$, respectively, are $\gamma'$ and $c'$, respectively, these two conditions yield
\be
c'\gamma'={R_3'-R_3+(R_3+R'_3)V_3(a') \gamma\over R_3'+R_3+(R'_3-R_3)V_3(a')\gamma}
={R'_3-R'_2+(R'_3+R'_2)(\beta+\beta')V_{2'}(a')\over R'_3+R'_2+(R'_3-R'_2)(\beta+\beta')V_{2'}(a')}\ ,\label{c'gamma'}
\ee
where we used $U_{2'}(\beta)=\beta+\beta'$.
Using the two-shock patchwise coordinates \eqref{eq:BTZpatchsolution}, it is easy to find that 
\be
V_3(a')\gamma={R_3-R_2+(R_3+R_2)\beta A\over R_3+R_2+(R_3-R_2)\beta A}
\ee
and
\be
(\beta+\beta')V_{2'}(a')={R_2'-R_2+(R_2'+R_2)\beta A\over R_2+R_2'+(R'_2-R_2)\beta A}\ .
\ee
Then one can check that \eqref{c'gamma'} reduces to
\be
c'\gamma'={R_3'-R_2+(R'_3+R_2)\beta A\over R_3'+R_2+(R'_3-R_2)\beta A}={(R_3'-R_2)+(R_3'+R_2)\beta A\over (R_3'+R_2)+(R_3'-R_2)\beta A}\ .
\ee
Meanwhile, multiplying \eqref{eq:v3prime} and \eqref{eq:u3prime} yields
\begin{align}
V_{3'}'(V_3(a'))U_{3'}'(U_{2'}(\beta))V_3(a') U_{2'}(\beta){R_3R'_2\over (R_3')^2} = c'\gamma' \ .
\end{align}
Calculating $U_{3'}'(U_{2'}(\beta))$ and $V_{3'}'(V_3(a'))$ from  \eqref{2PIII'} and \eqref{2PII'III'}, a little manipulations yield
\begin{align}
&{\left(R_3'-R_2+(R'_3+R_2)\beta A\right)^2\left(R_3'+R_2+(R'_3-R_2)\beta A\right)^2\over \left(R_3-R_2+(R_3+R_2)\beta A\right)^2\left(R_3+R_2+(R_3-R_2)\beta A\right)^2}(R_3^2-R^2)\nn\\
&\hspace{1cm}=(a+\alpha)(\beta+\beta')^2{R(R_2+R)^2\over 16\beta R_2'^2R_2^3}\left(R_2'+R_2+(R_2'-R_2)\beta A\right)^4\ .
\end{align}
Finally, by using the two-shock consistency condition \eqref{BHmassredTraversable}, we obtain 
\begin{align}
 &\hspace{-1cm}{\left(R_3'-R_2+(R'_3+R_2)\beta A\right)\left(R_3'+R_2+(R'_3-R_2)\beta A\right)\over \left(R_3-R_2+(R_3+R_2)\beta A\right)\left(R_3+R_2+(R_3-R_2)\beta A\right)}\nn\\
 &\hspace{1cm}={\beta+\beta'\over 2\beta}{\left(R_2'+R_2+(R_2'-R_2)\beta A\right)^2\over 2R_2'R_2}\ .\label{R3prime}
 \end{align}
 The consistency conditions \eqref{R3prime}, \eqref{R2prime} and \eqref{Arelation} can readily be generalized to the generic $i$-th P-shock:
\begin{align}
&\hspace{-1cm}{\left(R^{(i+1)}_3-R^{(i)}_2+(R^{(i+1)}_3+R^{(i)}_2)\beta_{i} A_{i+1}\right)\left(R^{(i+1)}_3+R^{(i)}_2+(R^{(i+1)}_3-R^{(i)}_2)\beta_{i} A_{i+1}\right)\over \left(R^{(i)}_3-R^{(i)}_2+(R^{(i)}_3+R^{(i)}_2)\beta_{i} A_{i+1}\right)\left(R^{(i)}_3+R^{(i)}_2+(R^{(i)}_3-R^{(i)}_2)\beta_{i} A_{i+1}\right)}\nn\\
&={\beta_{i}+\beta_{i+1}\over 2\beta_{i}}{\left(R^{(i+1)}_2+R^{(i)}_2+(R^{(i+1)}_2-R^{(i)}_2)\beta_{i} A_{i+1}\right)^2\over 2R^{(i+1)}_2R^{(i)}_2}\ ,
\end{align}
\be
{R^{(i+1)}_2\left(R^{(i+1)}_2-R^{(i)}_2+(R^{(i+1)}_2+R^{(i)}_2)\beta_{i} A_{i+1}\right)\over R^{(i)}_2\left(R^{(i+1)}_2+R^{(i)}_2+(R^{(i+1)}_2-R^{(i)}_2)\beta_{i} A_{i+1}\right)}= A_{i+1}(\beta_{i}+\beta_{i+1})\ ,
\ee
and
\be
A_{i+1}R_2^{(i)}=V_2^{(i+1)}(a_{i+1})R_2^{(i+1)}\ ,
\ee
where 
\be
A_{i+1}=V_2^{(i)}(a_{i+1})\qquad\mbox{and}\qquad V_2^{(i+1)}(V)=V_2^{(i)}(V)-V_2^{(i)}(a_{i+1})+V_2^{(i+1)}(a_{i+1})\ .
\ee
The particular expressions in \eqref{A-recursion}, \eqref{R2-recursion} and \eqref{R3-recursion} can be obtained by massaging these equations.

\addcontentsline{toc}{section}{References}
\bibliography{bib}
\bibliographystyle{newutphys}

\end{document}